\documentclass[12pt,a4paper]{article}
\usepackage{xcolor}
\usepackage{jheppub}
\usepackage{epstopdf}
\usepackage{graphicx}
\usepackage{epsfig}
\usepackage{dcolumn}  
\usepackage{bm}    
\usepackage{amssymb} 
\usepackage{amsmath,bm}
\usepackage{amsfonts}    
\usepackage{slashed}  
\usepackage{youngtab}
\usepackage[mathscr]{euscript}
\usepackage{epsfig}
\usepackage{verbatim}
\newdimen\tableauside\tableauside=1.0ex
\newdimen\tableaurule\tableaurule=0.4pt
\newdimen\tableaustep
\def\phantomhrule#1{\hbox{\vbox to0pt{\hrule height\tableaurule width#1\vss}}}
\def\phantomvrule#1{\vbox{\hbox to0pt{\vrule width\tableaurule height#1\hss}}}
\def\sqr{\vbox{%
		\phantomhrule\tableaustep
		\hbox{\phantomvrule\tableaustep\kern\tableaustep\phantomvrule\tableaustep}%
		\hbox{\vbox{\phantomhrule\tableauside}\kern-\tableaurule}}}
\def\squares#1{\hbox{\count0=#1\noindent\loop\sqr
		\advance\count0 by-1 \ifnum\count0>0\repeat}}
\def\tableau#1{\vcenter{\offinterlineskip
		\tableaustep=\tableauside\advance\tableaustep by-\tableaurule
		\kern\normallineskip\hbox
		{\kern\normallineskip\vbox
			{\gettableau#1 0 }%
			\kern\normallineskip\kern\tableaurule}%
		\kern\normallineskip\kern\tableaurule}}
\def\gettableau#1 {\ifnum#1=0\let\next=\null\else
	\squares{#1}\let\next=\gettableau\fi\next}

\newcommand{\ysm}{\Yboxdim{8.5pt}\Ylinethick{.9pt}}
\def\tyng{\ysm\yng}

\newcommand{\be}{ \begin{equation}}
\newcommand{\ee}{\end{equation}}
\newcommand{\bea}[1]{\begin{eqnarray}\label{#1} }
\newcommand{\eea}{\end{eqnarray}}

\def\ZZZ{{\hskip-3pt\hbox{ Z\kern-1.6mm Z}}}
\def\zzz{{\hskip-3pt\hbox{ z\kern-1mm z}}}

\def\lm{\lambda}
\def\V#1#2{ {V^{(#1)}_{#2}} }

\def\one{{\hbox{ 1\kern-.8mm l}}}
\def\zero{{\hbox{ 0\kern-1.5mm 0}}}

\def\th{\theta}

\title{Twisted sectors from plane partitions}

\author{Shouvik Datta$^a$, Matthias R.\ Gaberdiel$^a$,  Wei Li$^b$  and Cheng Peng$^a$} 
\affiliation{$^a$ Institut f\"ur Theoretische Physik, ETH Zurich, \\
\hspace*{0.3cm}CH-8093 Z\"urich, Switzerland}
\affiliation{$^b$ Institute of Theoretical Physics, Chinese Academy of Science\\
\hspace*{0.3cm} 100190 Beijing, P.R.\ China}
\emailAdd{shouvik@itp.phys.ethz.ch, gaberdiel@itp.phys.ethz.ch, weili@itp.ac.cn, pengch@itp.phys.ethz.ch}

\abstract{Twisted sectors arise naturally in the bosonic higher spin CFTs at their free points, as well as in the
associated symmetric orbifolds.  We identify the coset representations of the twisted sector states using 
the description of   ${\cal W}_\infty$ representations in terms of plane partitions. We confirm these proposals
by a microscopic null-vector analysis, and by matching the excitation spectrum of these representations 
with the orbifold prediction.
}

\begin{document}

\maketitle

\makeatletter
\g@addto@macro\bfseries{\boldmath}
\makeatother
\section{Introduction}

Massless higher spin theories can be constructed consistently  on AdS backgrounds \cite{Vasiliev:2003ev}, and they are 
believed to describe a consistent subsector of string theory at the tensionless point \cite{Sundborg:2000wp,Witten,Mikhailov:2002bp}.
Via the AdS/CFT duality, the tensionless limit of string theory corresponds, on the dual field theory side, to the limit in which the 
CFT becomes free, and the higher spin degrees of freedom correspond to those of a vector-like CFT. The AdS/CFT
correspondence therefore predicts dualities between higher spin theories and vector-like CFTs, and
explicit examples of such relations were first proposed for the case of ${\rm AdS}_4 / {\rm CFT}_3$ in \cite{Klebanov:2002ja,Sezgin:2003pt},
and more recently in one dimension lower in \cite{Gaberdiel:2010pz}. These dualities provide a 
useful approach towards analysing string theory at a very symmetrical point in its moduli space where many of its 
underlying symmetries are unbroken. They may also allow one to prove the 
AdS/CFT correspondence since they are weak-weak dualities.

In order to utilize the duality for either of these purposes it is important to understand the embedding of the higher spin AdS/CFT
duality into the usual stringy AdS/CFT correspondence in detail. For the 4d/3d case, a proposal was made some time ago in 
\cite{Chang:2012kt}, while for the case of ${\rm AdS}_3 / {\rm CFT}_2$ a somewhat different picture emerged in 
\cite{Gaberdiel:2014cha}. The CFT duals of string theory on ${\rm AdS}_3\times {\rm S}^3 \times \mathbb{T}^4$ lie on the 
moduli space that contains the symmetric orbifold of $\mathbb{T}^4$, and the orbifold theory itself contains a vector-like 
CFT as a closed subsector. In turn, this
vector-like CFT was shown to emerge naturally in the CFT dual of the ${\cal N}=4$ version \cite{Gaberdiel:2013vva} of the
higher spin AdS/CFT duality. More specifically, the CFT dual of the higher spin theory is a subsector of the untwisted sector
of the symmetric orbifold, and the entire untwisted sector can be understood in terms of a vastly extended higher spin symmetry,
the so-called Higher Spin Square (HSS), as well as its scalar field excitations \cite{Gaberdiel:2015mra,Gaberdiel:2015wpo},
see also \cite{Baggio:2015jxa, Gaberdiel:2015uca} for related discussions. 
On the other hand, it has been more difficult to characterize  the twisted sector of the symmetric orbifold from a higher spin
perspective --- see however \cite{Jevicki:2015irq,Gaberdiel:2015wpo,Gaberdiel:2016xwo} for first steps in this direction. This is an important problem if we want to use the higher spin perspective for the description of string theory at its highly 
symmetrical tensionless point.
\smallskip

The present work revisits the original bosonic duality of \cite{Gaberdiel:2010pz} in order to analyse the 
relevant twisted sectors from a higher spin perspective. The bosonic theory is a useful toy model since it exhibits 
all the essential features of the supersymmetric version.
The identification of the twisted sectors in terms of coset representations was done before for the 
${\cal N}=2$ and ${\cal N}=4$ cases  \cite{Gaberdiel:2014vca,Gaberdiel:2014cha}, partially using the BPS condition 
as a guide, but the description in the bosonic case has so far not been worked out. In fact, the structure of the 
bosonic coset is somewhat different from that of the supersymmetric versions, and thus it was not clear how 
to generalize the results directly. 
\smallskip

In this paper we attack this problem using a new tool that was recently
discovered for the bosonic ${\cal W}_\infty$ algebras. A few years ago it was shown in a series of papers
\cite{FFJMM1,FFJMM2,FJMM1} that the representation theory of the quantum toroidal algebra of $\mathfrak{gl}_{1}$
can be described in terms of plane partitions, and that the associated characters are, up to an overall $q$-Pochhammer symbol,
identical to those of the 
bosonic ${\cal W}_{N,k}$ minimal models. More recently, Proch\'{a}zka \cite{Prochazka:2015deb} 
realized that this gives rise to a powerful method to analyse the bosonic ${\cal W}_\infty$ representations. 
In particular, he showed that the triality symmetry 
of the ${\cal W}_\infty$ algebra, which played an important role in establishing the duality \cite{Gaberdiel:2012ku}, is inherent and manifest in the plane partition description. Finally, since quantum toroidal algebras are isomorphic to their corresponding affine Yangian algebras 
(after suitable completion) \cite{GTL,Tsym}, plane partitions also describe the representation theory of a 
Yangian algebra. On the other hand, Yangian algebras are one of the hallmarks of integrability, and hence
this viewpoint may help us establish the precise connection between higher spin theories and
integrable field theories proposed for AdS$_3$ in \cite{OhlssonSax:2011ms,Sax:2012jv,Borsato:2015mma} (see \cite{Sfondrini:2014via} for a review).
\smallskip

The ${\cal N}=4$ construction of \cite{Gaberdiel:2014cha} relates the Wolf space cosets to the symmetric
orbifold (and hence to string theory) for the case where the cosets can be described in terms of free fields ($\lambda=0$). 
The free field constructions of the bosonic ${\cal W}_\infty[\lambda]$ algebras arise for $\lambda=0$ and $\lambda=1$, where they can be 
realized in terms of free fermions and bosons, respectively
\cite{Bergshoeff:1989ns,Bergshoeff:1990yd,Depireux:1990df,Bakas:1990ry,Gaberdiel:2013jpa}.
For the case of $\lambda=0$, the free field realization of the coset model was already made fairly explicit in \cite{Gaberdiel:2011aa}, 
where the $k\rightarrow \infty$  (i.e.\ $\lambda\rightarrow 0$) limit  
was described as a continuous orbifold. 
However, the precise description of the twisted sectors was not understood at the time ---  in general, the twisted 
sector of the orbifold is not directly accessible, even if one has a good understanding
of the untwisted sector. In this paper we find the coset description of the twisted sectors both at the free fermion 
($\lambda=0$) and the free boson point ($\lambda=1$). 
It is important 
to have both of these cases simultaneously under control since the extended higher spin symmetry algebra that is believed to arise in string 
theory, the Higher Spin Square (HSS), is in some sense a combination of both of these constructions \cite{Gaberdiel:2015mra}. 
The main technical
advance of our analysis is the systematic use of the plane partition viewpoint advocated in \cite{Prochazka:2015deb},
which enabled us to find the correct twisted sector representations. We also test our proposals using the techniques
of \cite{Gaberdiel:2014vca}.
\medskip

The paper is organized as follows. We begin in Section~\ref{sec:2} with a discussion of the bosonic theory corresponding to 
$\lambda=1$, i.e.\ $N\rightarrow\infty$ at fixed $k$. We first find closed form expressions for the wedge
characters of the twisted sectors, and then use the plane partition viewpoint to propose the form of the corresponding
coset representations. This proposal is then tested in some detail: in Sections~2.3 the null-vector structure of the 
corresponding ${\rm hs}[\lambda]$ representations are studied from a microscopic viewpoint, i.e.\  by calculating the 
relevant Kac determinants, and in Section~2.4 the conformal dimension and excitation spectrum is computed in the 
coset and found to agree with the orbifold predictions; this also fixes the precise identification with the coset representations. 
In Section~\ref{sec:3} the corresponding analysis is performed
for the fermionic theory corresponding to $\lambda=0$. We end with a discussion on future directions in 
Section~\ref{sec:4}. There are three
appendices where aspects of the null-vector analysis (Appendix~\ref{app:level1}), the determination of the higher spin charges using the 
Drinfeld-Sokolov approach (Appendix~\ref{app:B}), and combinatorial identities that arise in the plane partition analysis
(Appendix~\ref{app:C}) are explained in more detail.

\section{The twisted sector in the free boson description}
\label{sec:2}

We are interested in the cosets 
\be\label{coset}
\frac{\mathfrak{su}(N)_k \oplus \mathfrak{su}(N)_1}{\mathfrak{su}(N)_{k+1}} \ , 
\ee
and we shall mainly be considering the 't~Hooft limit, where we take $N$ and $k$
to infinity, while keeping the ratio
\be\label{tHooft}
\lambda = \frac{N}{N+k} \
\ee
fixed. The case where we take $N\rightarrow\infty$ first then corresponds to the theory at $\lambda=1$. 
This limit theory can be described by a free boson 
construction, see e.g. \cite{Bakas:1990ry,Gaberdiel:2013jpa}. More specifically, for $k$ complex
bosons $\phi^i$ and $\bar\phi^i$ that transform in the fundamental and anti-fundamental
representation of ${\rm U}(k)$, respectively, we consider the chiral ${\rm U}(k)$ singlets that are of the form
\be
W^s(z) = m(s)  \, \sum_{l=1}^{s-1}  \frac{(-1)^l}{(s-1)} {s-1 \choose l} { s-1 \choose s-l} 
\partial^l\phi^j\, \partial^{s-l}\bar\phi^j \ ,
\ee
where $m(s)$ is an $s$-dependent normalization constant.
These currents then generate the ${\cal W}_{\infty}[1]$ algebra with $c=2k$. 
Formally, the $\lambda=1$ theory can therefore be thought of as a continuous orbifold, where we divide
the free boson theory by the orbifold group ${\rm U}(k)$, see \cite{Gaberdiel:2011aa}. 
The theory should then also contain twisted sectors where the different complex bosons 
are twisted. 

Note that some of these twisted sectors also appear in the symmetric orbifold 
where we divide the theory by the symmetric group $S_{k+1}\subset {\rm U}(k)$,
under which the above currents are also invariant. (The full chiral algebra of the
symmetric orbifold is then much bigger --- it gives rise to the stringy extension
of the ${\cal W}_{\infty}[1]$ algebra that is associated to the so-called Higher Spin
Square \cite{Gaberdiel:2015mra}.) The analysis of the twisted sectors is therefore
in particular relevant for the stringy embedding of the bosonic duality of \cite{Gaberdiel:2010pz}. 

In general, the coset interpretation of the twisted sectors cannot be deduced directly
from the identification in the untwisted sector, and we shall therefore proceed 
indirectly. 
Let us first concentrate on the case where only one complex boson is twisted by $\nu$ with
$0<\nu<1$; the twisted sector is then generated by 
\begin{equation}
\alpha_{n-\nu} \qquad \text{and}\qquad \bar{\alpha}_{n+\nu} \qquad \qquad \textrm{with} \quad n\in \mathbb{Z} \ .
\end{equation}
The wedge character of the corresponding representation is given by 
\cite{Gaberdiel:2015wpo}
\be\label{twgen}
\chi_{[\nu]} (q,y) = q^h \prod_{n=1}^\infty (1-y\,q^{n-1+\nu})^{-1} (1-y^{-1}\,q^{n-\nu})^{-1}  \ ,
\ee
where the conformal dimension is $h = \frac{1}{2} \nu (1-\nu)$. In the full orbifold theory,
this chiral representation comes together with a corresponding anti-chiral representation,
and on the full space (involving both chiral and anti-chiral twisted states) the invariance 
under the orbifold group is to be imposed. In particular, not just those states survive
this orbifold projection that are separately invariant under the orbifold action; instead the 
correct condition is that the left-moving states transform in the conjugate representation
to that of the right-moving states. The powers of $y$ keep track of the action under
the cyclic group corresponding to the twist itself, and hence the states corresponding
to a given fixed power of $y$ correspond to different representations of the 
${\rm hs}[1]$ algebra, 
\begin{equation}\label{wedgen}
\chi_{[\nu]}^{(\ell)} (q) \equiv \left. \chi_{[\nu]} (q,y)   \right|_{y^\ell} \ . 
\end{equation}

In the following we shall identify the coset representations that describe these twisted
sector states. We shall first use character considerations to make a proposal for the 
corresponding coset representation, see Sections~\ref{sec:char} and \ref{sec:char1}. 
In Section~\ref{sec:null} we study the representation corresponding to $\ell=0$ using the commutation relations of the
${\rm hs}[\lambda]$ algebra, and confirm in particular, that the representation we have identified 
in Section~\ref{sec:char} leads to the correct eigenvalues for arbitrary $\lambda$. (As will become
clear in the following, these representations can also be defined for general $\lambda$;
however, we do not have a direct interpretation in terms of an orbifold unless $\lambda=1$.) Finally, in 
Section~\ref{sec:exc}, we confirm that the representations have the correct ground state 
conformal dimension and excitation spectrum.

\subsection{Wedge characters and their combinatorial interpretation}\label{sec:char}

In order to use the characters for determining the corresponding coset representations, we first compute the wedge characters defined in  
eq.~(\ref{wedgen}). In particular, we want to find closed form expressions whose combinatorial interpretation  can help us identify 
their corresponding plane partition configurations.
\medskip

There are two factors in the wedge character (\ref{twgen}), corresponding to modes associated to 
$\phi$ and $\bar{\phi}$, respectively.  
Both of them resemble the refined version of the generating function of  partition numbers defined by 
\begin{equation}\label{partitionZy}
Z(q,y) = \prod_{n=1}^\infty \frac{1}{(1-y\,q^n)} =\sum_{m=0}^\infty y^m \,\sum_{n=m}^\infty p(n,m)\,q^n \ ,
\end{equation}
where $p(n,m)$ counts the number of Young diagrams which have $n$ boxes and whose 
height is $m$. Summing over all Young diagrams with the same fixed height $m$ we have 
\begin{equation}\label{pnmG}
 \sum_{n=m}^{\infty} p(n,m) \,q^n =q^m \prod^m_{n=1}\frac{1}{(1-q^n)} \ . 
\end{equation}
Expanding both factors in the wedge character $\chi_{[\nu]}(q,y)$ as in (\ref{partitionZy}) and (\ref{pnmG}) and collecting the coefficient of 
$y^{\ell}$ term, we obtain the expression for the wedge characters 
\begin{equation}\label{pnmlG1}
\chi^{(\ell)}_{[\nu]} (q) = q^{h+\delta h(\ell,\nu)}\,
\sum_{m=0}^\infty q^m \,\prod^{m+|\ell|}_{n=1}\frac{1}{(1-q^{n})} \,\prod^{m}_{n=1}\frac{1}{(1-q^{n})} \ ,
\end{equation}
where 
\be\label{deltah}
\delta h(\ell,\nu)= \left\{ 
\begin{array}{cl} 
\ell\, \nu & \qquad \qquad\hbox{$\ell\geq 0$} \\
\ell \,(\nu-1) & \qquad \qquad \hbox{$\ell<0$ .}
\end{array} \right.
\ee
corresponds to the excitation spectrum, and $\ell$ enters the combinatorial part of the character only as $|\ell|$. 
The explicit $q$-expansions of the first few values of $|\ell|$ are 
\begin{equation}\label{chiqexp}
\begin{aligned}
\chi_{[\nu]}^{(0)} (q)&= q^h \bigl(1+q+ 3q^2+ 6q^3+ 12q^4+ 21q^5+ 38q^6+63 q^7+\cdots \bigr)\\
\chi_{[\nu]}^{(1)} (q)&= q^{h+\nu} \bigl(1+2q+ 4q^2+ 8q^3+ 15q^4+ 27q^5+ 47q^6+79 q^7+\cdots\bigr)\\
\chi_{[\nu]}^{(2)} (q)&= q^{h+2\nu} \bigl( 1+2q+ 5q^2+ 9q^3+ 18q^4+ 31q^5+ 55q^6+91 q^7+\cdots\bigr)\\
\chi_{[\nu]}^{(3)} (q)&= q^{h+3\nu} \bigl(1+2q+ 5q^2+ 10q^3+ 19q^4+ 34q^5+ 60q^6+100 q^7+\cdots \bigr)\ .
\end{aligned}
\end{equation}
For the $\ell=0$ representation, there is one descendant at level one, while for $\ell \neq 0$, the 
representation has two descendants at level one. This property is also directly visible in 
(\ref{pnmlG1}) since the first product only contributes a state at level $1$ if $|\ell|>0$. 
\medskip

\begin{figure}[!h]
	\centering
	\begin{tabular}{c}
		\includegraphics[width=.5\textwidth]{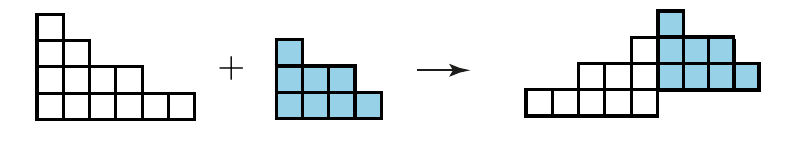} 
	\end{tabular}
	\caption{\small Gluing two Young diagrams along their first columns: the difference in height is $\ell=1$ and the total number of visible boxes after gluing is  $n=17$.}
	\label{overlap-young}
\end{figure}

Analogous to the counting in eq.~(\ref{pnmG}), the wedge character $\chi_{[\nu]}^{(\ell)} (q)$ has a combinatorial interpretation 
\be\label{p2ndef}
\chi^{(\ell)}_{[\nu]} (q) = q^{h +\delta h(\ell,\nu)}\, \sum_{n=0}^{\infty} p_2(n,\ell)\, q^n \ , 
\ee
where $p_2(n,\ell)$ counts pairs of Young diagrams $\Gamma^\pm$ whose height difference 
is $\ell$, i.e.\ $c^+_1 - c^-_1 = \ell$. Here $c^\pm_i$ are the number of boxes in the $i$'th row of $\Gamma^\pm$, and 
$n$ is the combined number of boxes in the two Young diagrams except that, for the first column of each
of the two Young diagrams, only the boxes of the shorter diagram are counted, i.e.  
\begin{equation}\label{funnyn}
n = \min (c^+_1, c^-_1 ) + \sum_{i=2}^{} (c^+_i  + c^-_i )   \ .
\end{equation}
The reason for this unusual condition is that in the first factor of $\chi_{[\nu]} (q,y)$ 
the prefactor is $y\, q^{-1}$ (instead of $y$). 
A useful way to visualise this configuration is by first raising the shorter Young diagram (with height $m$) to the same height as that of the 
taller one (with height $m+|\ell|$), then gluing the two Young diagrams together along their first columns, and finally removing the $|\ell|$
boxes in the first column of the taller Young diagram that are not covered by the shorter diagram, see Figure~\ref{overlap-young}.

There is an alternative formula for the wedge character that makes 
the connection to the original complex boson more transparent,
\begin{equation}
\begin{aligned}\label{twistedbosonwedge}
\chi_{[\nu]}^{(\ell)} (q) =q^{h+\delta h(\ell,\nu)} \left(\sum_{m=0} ^\infty (-1)^m \,
 q^{\,{\sum_{k=|\ell|+1}^{|\ell|+m} k}}
\right)
\cdot  \prod_{n=1}^\infty  \frac{1}{(1-q^n)^2} \ .
\end{aligned}
\end{equation}
This can be obtained from eq.~(\ref{p2ndef}), using the combinatorial identity
\begin{equation}\label{p2rel}
p_2(n,\ell)=\sum_{m=0}^\infty (-1)^m\,p_2 \Bigl(n-\sum^{|\ell|+m}_{k=|\ell|+1}k \Bigr) \ ,
\end{equation}
where $p_2(n)$ counts pairs of Young diagrams whose total number of boxes is $n$ 
via the generating function of the complex boson
\begin{equation}
\begin{aligned}
Z_{{\rm cplx}\, {\rm bos}}(q)& = \prod_{n=1}^{\infty} \frac{1}{(1-q^n)^2} = \sum_{n=0}^{\infty} p_2(n)\, q^n \\
&= 1+2 q + 5 q^2 + 10 q^3 + 20 q^4 + 36 q^5 + 65 q^6 +110 q^7+ \cdots \ . 
\end{aligned}
\end{equation}
A proof of (\ref{p2rel}) for $\ell=0$ is given in \cite{combinatorics}; we give the generalization of the proof 
to arbitrary $\ell$ in Appendix~\ref{app:C}. In this formula the property that the wedge
character only has a single descendant at level one for $\ell=0$ comes from the fact that the 
first factor starts with $1-q^{|\ell|+1}+\cdots$, i.e.\ it removes a state at level one for $\ell=0$ but
not otherwise.
\smallskip

As an aside we also mention that the asymptotics of the two-partition function $p_2(n)$ is \cite{Keane:2002} 
\begin{equation}\label{p2asymp}
p_2(n)\sim \frac{3^{\frac{1}{4}}}{12  } \cdot n^{-\frac{5}{4}} \cdot \exp{\left(\tfrac{2}{\sqrt{3}}\, \pi \, \sqrt{n}\right)} \ , 
\end{equation} 
whereas $p_2(n,\ell)$ with $\ell=0,1$ --- we expect that the same is true for general $\ell$ --- grows half as fast:
\begin{equation}\label{p2asymp}
p_2(n,\ell)\sim \frac{1}{2}\, p_2(n) \ .
\end{equation} 
Neither of them grows much faster than the ordinary partition  numbers \cite{HardyRamanujan}
\be\label{pasymp}
p(n)\sim \frac{1}{4\sqrt{3}  } \cdot \frac{1}{n} \cdot \exp{\left(\sqrt{\tfrac{2}{3}}\, \pi \, \sqrt{n}\right)} \ , 
\ee
since the exponential in both cases is proportional to $\sqrt{n}$.

\subsection{Plane partition viewpoint}\label{sec:char1}

We have seen in the previous section that the wedge characters of the twisted sector eq.~(\ref{wedgen}) 
can be interpreted as counting the configurations of two Young diagrams that are glued together along their first 
columns. This viewpoint now allows us to determine the corresponding coset representation, using the 
description of the \linebreak ${\cal W}_\infty$ representations in terms of plane partitions, which we shall now review.
\medskip

Just as the partition of $n$ counts the number of ways of drawing Young diagrams with $n$ boxes, the plane 
partition of $n$ counts the number of ways of stacking $n$ boxes in the corner of a room (such that the number 
of boxes is non-increasing along all three directions, i.e.\  the projections onto the $xy$, $yz$, $zx$ planes all have 
the shape of a Young diagram).
The generating function of the plane partitions is the MacMahon function 
\begin{equation}\label{MacMahon}
\begin{aligned}
 M(q)
  &\equiv\prod^{\infty}_{s=1} \prod^{\infty}_{n=s} \frac{1}{(1-q^n)}=  \prod^{\infty}_{n=1}\frac{1}{(1-q^n)^n}  
   = \sum_{n=0}^{\infty} M(n)\, q^n \\
  & = 1+q + 3 q^2 + 6 q^3 + 13 q^4 + 24 q^5 + 48 q^6 +86 q^7+ \cdots \ . 
\end{aligned}
\end{equation}
From the definition of the MacMahon function, it is immediate that it is identical to the vacuum 
character of the ${\cal W}_{1+\infty}$ algebra --- for each spin $s=1,2,\ldots$, the modes that
contribute to the vacuum Verma module are those with $n=-s,-s-1,\ldots$. It has the asymptotic behaviour  \cite{Wright1931}
\begin{equation}\label{asympP}
M(n) \sim \frac{\zeta(3)^{\frac{7}{36}}2^{\frac{25}{36}}}{\sqrt{12 \pi} } \, e^{\zeta'(-1)}  \cdot n^{-\frac{25}{36}} 
\cdot \exp{\left( \frac{3 \zeta(3)^{\frac{1}{3}} }{2^{\frac{2}{3}}} n^{\frac{2}{3}} \right)} \ . 
\end{equation}
Because of the `$n^{\frac{2}{3}}$' in the exponent, it grows much faster than the 
ordinary partition (\ref{pasymp}) or the two-partition (\ref{p2asymp}), whose exponents
are propotional to $\sqrt{n}$.

What is more interesting is that we can also consider the set of plane partitions that share a given 
asymptotic behaviour described by $(\Lambda_x, \Lambda_y, \Lambda_z)$, where $\Lambda_a$ 
with $a=x,y,z$ is the Young diagram to which the plane partition asymptotes in the limit $a \rightarrow\infty$. 
For a given asymptotic $(\Lambda_x, \Lambda_y, \Lambda_z)$,  there exists 
a unique plane partition configuration that has the least number of boxes --- let's call it the minimal configuration 
with this boundary condition. The character of plane partitions 
$\mathcal{N}_{\Lambda_x, \Lambda_y, \Lambda_z}(q)$ 
counts the number of ways of stacking boxes starting from this minimal configuration. 

When all three asymptotics are trivial, the character 
$\mathcal{N}^{\textrm{plane}}_{\Lambda_x, \Lambda_y, \Lambda_z}(q)$ reduces to the MacMahon function --- 
the vacuum character of the $\mathcal{W}_{1+\infty}$ algebra. When at least one of the three Young diagrams 
is trivial --- without loss of generality we may take $\Lambda_z=0$ --- the generating function of the plane partition 
reproduces  precisely the $\mathcal{W}_{1+\infty}$ character for the coset representation 
$(
\Lambda_x, \Lambda_y
)$ \cite{FFJMM1,FFJMM2,FJMM1,Prochazka:2015deb} 
\begin{equation}\label{PPasWirrep}
\mathcal{N}^{\textrm{plane}}_{\Lambda_x, \Lambda_y, 0}(q) 
=\chi^{\mathcal{W}_{1+\infty}}_{(\Lambda_x; \Lambda_y)}(q) \ . 
\end{equation}
Here the representations of the coset (\ref{coset}) are labelled by the pairs $(\Lambda_+;\Lambda_-)$,
where $\Lambda_+$ denotes a representation of $\mathfrak{su}(N)_k$ in the numerator, while
$\Lambda_-$ denotes a representation of $\mathfrak{su}(N)_{k+1}$ in the denominator.\footnote{The plane 
partitions describe representations of ${\cal W}_{1+\infty}$, while the coset defines 
a ${\cal W}_\infty$ algebra in the large $N$ limit. These two algebras are, however, closely related since one
can always decouple the spin $1$ current, and hence ${\cal W}_{1+\infty} \cong \mathfrak{u}(1) \oplus {\cal W}_\infty$.
Therefore every plane partition asymptotics gives rise to a representation of ${\cal W}_\infty$. Furthermore, 
their wedge representations coincide since the only wedge mode of the spin $1$ current is a central zero mode.}
(The representation
of the $\mathfrak{su}(N)_1$ is then uniquely fixed by the selection rules.) The interpretation
of the representations with three non-trivial Young diagrams is not yet entirely clear, although
\cite{Prochazka:2015deb} has argued that the exchange of the three asymptotic directions reflects
precisely the `triality' symmetry of \cite{Gaberdiel:2012ku}.
\medskip

We will now use the map between the generating function of the plane partitions and the 
$\mathcal{W}_{1+\infty}$ characters  (\ref{PPasWirrep}) to identify the coset representations 
which  correspond to the twisted sector states. First of all, the configurations of two glued Young diagrams 
that are counted by $p_2(n,\ell)$, see eq.~(\ref{p2ndef}) above, 
 can be described, in the plane partition language, as the plane
partitions with a pit dug at $(x,y)=(2,2)$ \cite{BFM}. Here, the 
presence of a `pit'  means that one cannot place a box at  that position. But since a plane partition has to 
give Young diagrams upon the projection along all three directions, a `pit' at $(x,y)=(2,2)$ means that 
we cannot place any box at a position with $x\geq 2$ or $y\geq 2$. The plane partitions with this `pit' 
condition therefore reduce to a pair of Young diagrams that are glued along their first columns, where 
the two Young diagrams sit in the $zx$ and $yz$ planes, respectively, and the shared first 
column is along the $z$-direction.

\begin{figure}[!h]
	\centering
	\begin{tabular}{c}
	         \includegraphics[width=.3\textwidth]{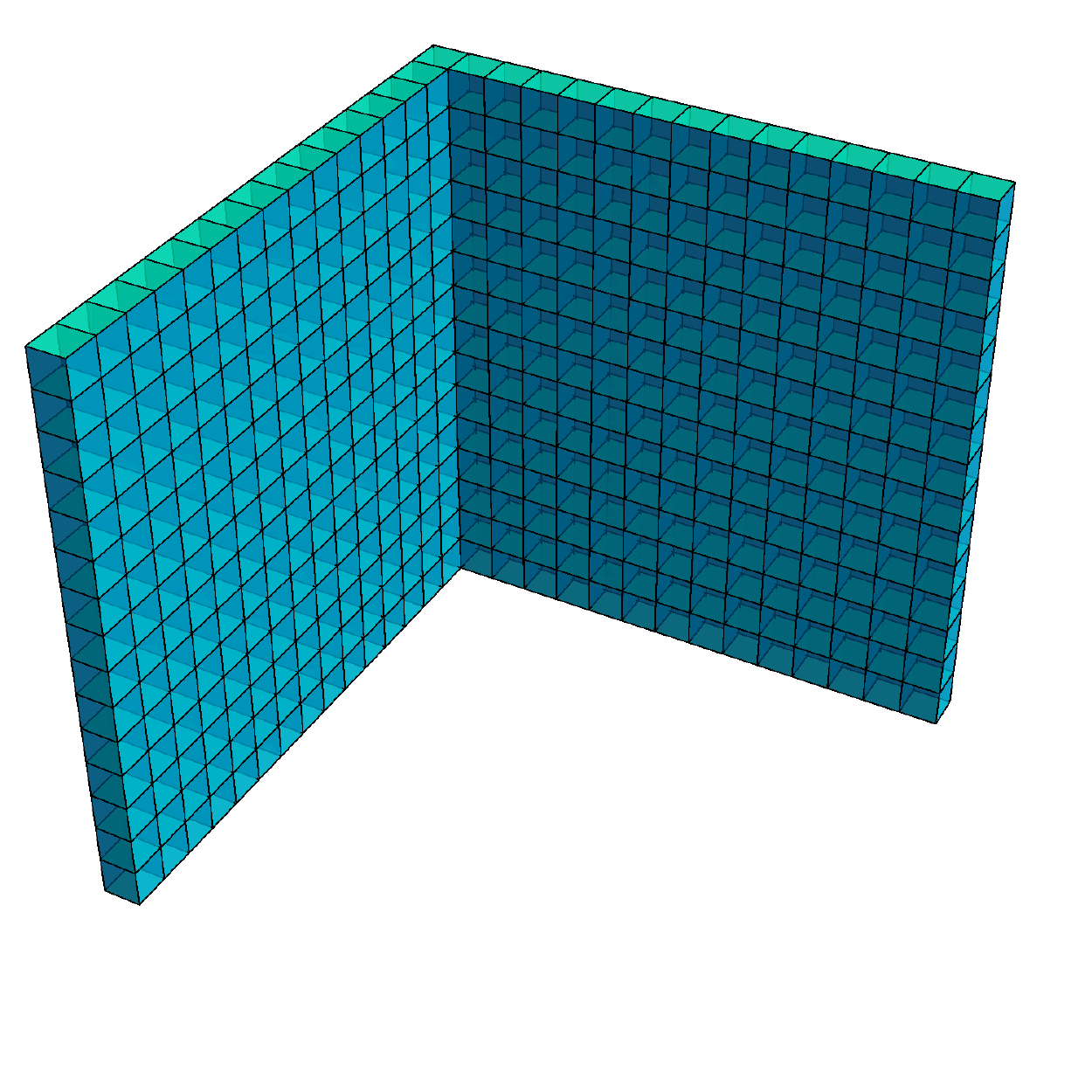} \qquad \qquad 
		\includegraphics[width=.3\textwidth]{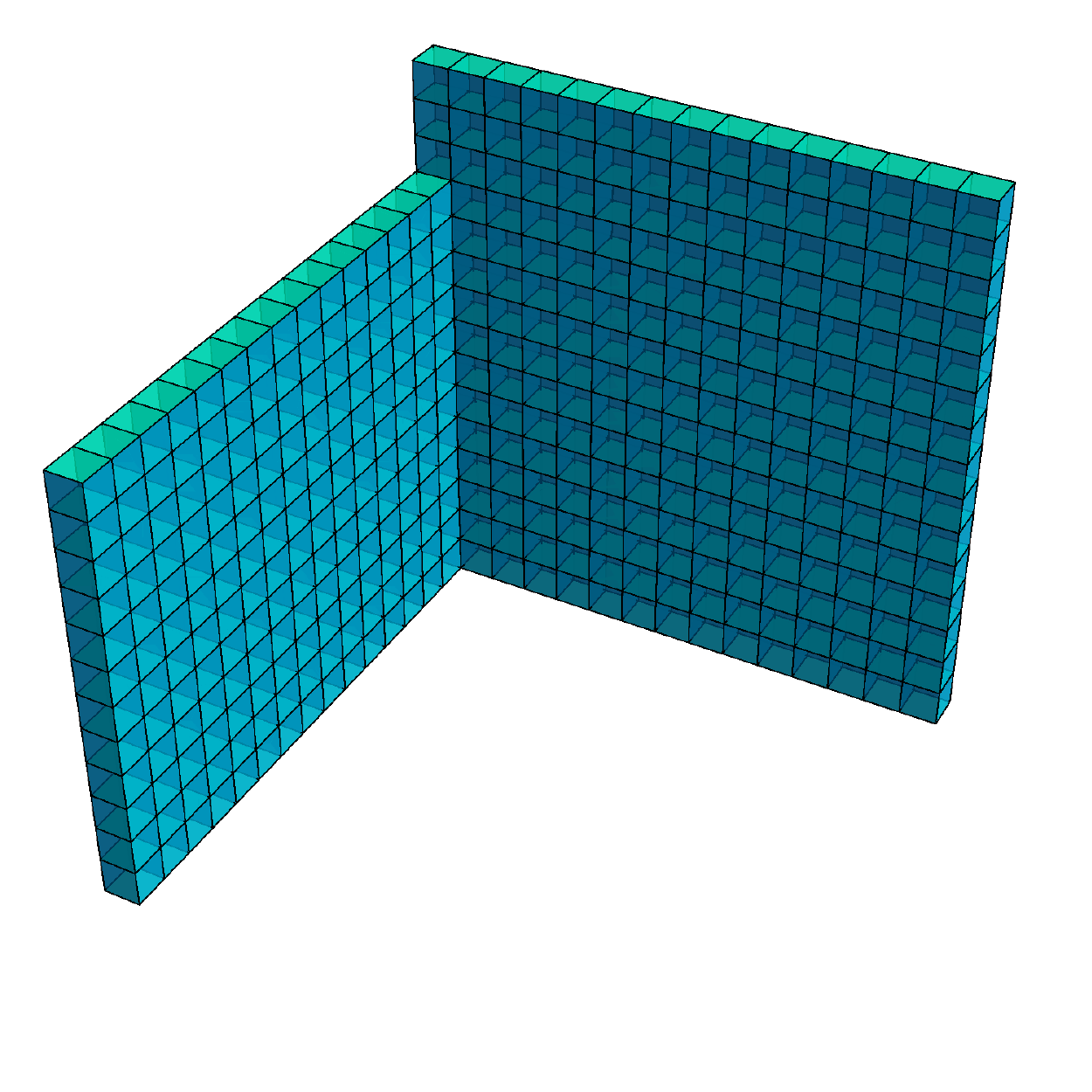}  \\ 
		\vspace{-1.5cm} (a) \hspace{5.5cm} (b) \vspace{1.5cm}
	\end{tabular}
	\caption{The minimal configuration of plane partitions that describes (a) the terms with 
	$y^0$ in eq.~(\ref{twgen}); and (b)  the terms with $y^3$ in eq.~(\ref{twgen}).
	The corresponding coset representations are described by (\ref{bostwist1}).}
	\label{fig1}
\end{figure}

Next we recall that eq.~(\ref{p2ndef}), i.e.\  the plane partition with the `pit' condition, 
only counts the wedge character. 
The full character is obtained by multiplying the wedge character with the vacuum character. Since the first 
one is given by the `pit' partition function, whereas the second equals the MacMahon function --- the 
plane partition starting from an empty corner --- the full coset character is then described by 
the window sill configuration of Figure~\ref{fig1} in the limit in which the height of the walls are taken to 
infinity. Indeed, in this limit, there is a natural separation between 
the configurations that involve boxes being stacked on the `floor' --- these are counted
by the MacMahon function, and hence describe the contribution of the ${\cal W}_{1+\infty}$ modes outside
the wedge --- and those that are stacked on the high `window-sill', and which are counted
by the plane partition with pit at $(x,y)=(2,2)$.

It remains to relate the height $b$ of the window sill to the twist of the corresponding bosonic representation. 
By comparing conformal dimensions, see Section~\ref{sec:exc}, we find that the relevant coset representations are
\be\label{bostwist1}
(\Lambda_+;\Lambda_-) = \bigl( [0^{b-1},1,0,\ldots,0]; [0^{b+\ell-1},1,0,\ldots,0] \bigr) \ ,
\ee
where 
\begin{equation}
b=\nu N \ .
\end{equation}

\begin{figure}[!h]
	\centering
	\begin{tabular}{c}
	         \includegraphics[width=.3\textwidth]{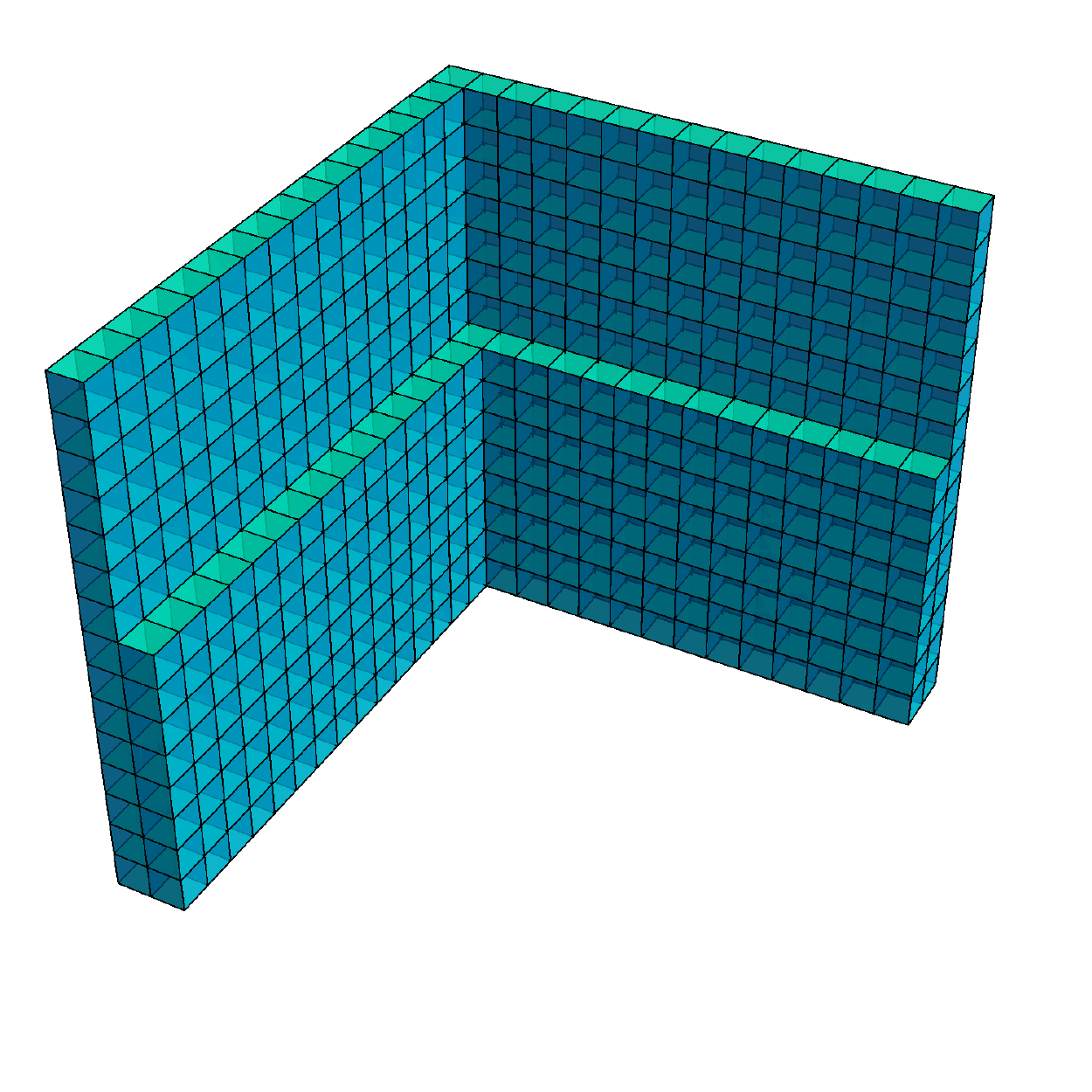} \qquad \qquad 
		\includegraphics[width=.3\textwidth]{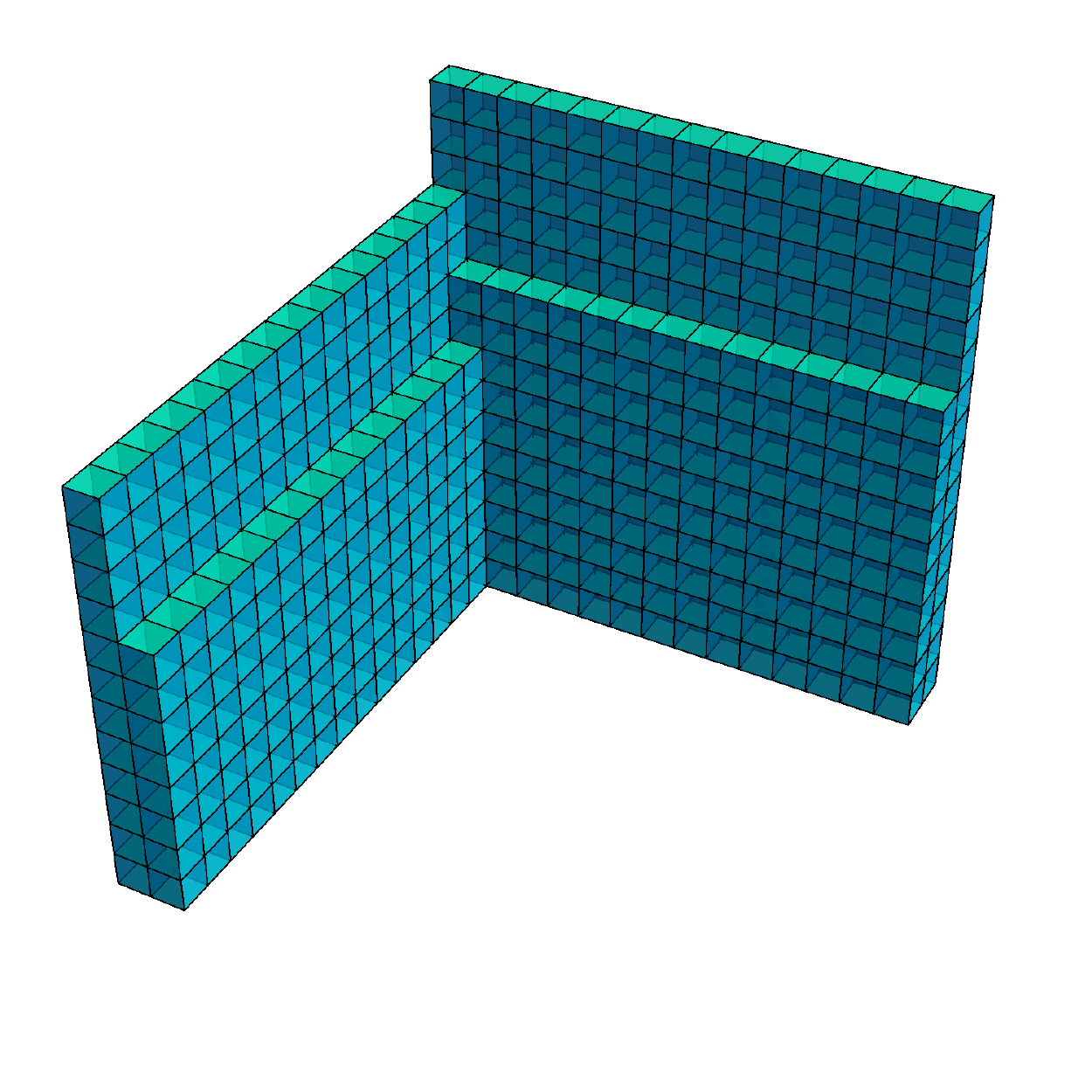}  \\ 
		\vspace{-1.5cm} (a) \hspace{5.5cm} (b) \vspace{1.5cm}
	\end{tabular}
	\caption{The  minimal configuration of plane partitions that captures the situation where two complex bosons are twisted. Case (a) describes
	the ground state representation, while (b) describes the representation appearing at some excited level.}
	\label{fig2}
\end{figure}

Furthermore, the case where more than one boson is twisted is described by putting the relevant window-sills
together, see Figure~\ref{fig2} for an example where two bosons are twisted. The generalization to the situation
where some or all bosons are twisted is then straightforward, where the height of each window-sill 
should be identified with $\nu_i N$, where $\nu_i$ is the twist parameter of the corresponding boson.
In particular, there is then a natural separation for the different box configurations into the boxes on the `floor' --- again
these configurations describe the contributions of the outside-the-wedge modes --- and the boxes stacked on the individual 
window-sills.\footnote{As long as the twists $\nu_i$ are pairwise disjoint, the height-differences also scale with $N$, and
hence the different window-sill contributions `decouple'. On the other hand, if two twists agree precisely, actually fewer states 
in the twisted sector survive since then the centralizer includes in particular the exchange of the two bosons with
the same twist --- this is accounted for by the counting of boxes on a window-sill of width $2$, etc.}
The wedge character in the multi-twist case is therefore just the product of the individual wedge characters
(\ref{twistedbosonwedge}).

These considerations therefore suggest that the coset representation $(\Lambda_+; \Lambda_-)$ that corresponds 
to the ground state of the multi-twisted sector
associated to the twist $(\nu_1,\ldots,\nu_k)$ with $\nu_1\geq \nu_2\geq \cdots \geq \nu_k$ 
is of the form $(\Lambda_+;\Lambda_-)$, where 
\be\label{twisgen}
\Lambda_+=\Lambda_-=\Lambda \qquad \textrm{with}\quad c_i = \nu_i N \,\,\,\hbox{boxes in the $i$'th column.}
\ee
Furthermore, the various twisted excitations (corresponding to the non-trivial powers of $y_i^{\ell_i}$) 
are described by the coset representations for which a finite number of boxes 
(corresponding to the number of twisted excitation modes) are added to or removed from
$\Lambda_-$ (but not  $\Lambda_+$). Namely, the column heights of $\Lambda_+$ and $\Lambda_-$ are 
\begin{equation}\label{twistexc}
c_i^+ = \nu_i N \ , \qquad c_i^- = \nu_i N + \ell_i  \ ,
\end{equation}
where $\ell_i$ is finite (and does not scale with $N$). 
These predictions will be tested below, see Section~\ref{sec:exc}, 
following the techniques of \cite{Gaberdiel:2014vca}.

\subsection{The null-vector analysis}\label{sec:null}

One of the key results of  the previous two subsections is that the coset representation corresponding 
to the ground state of the single-twist sector is of the form 
(\ref{bostwist1}), i.e.\  given by two totally anti-symmetric Young diagrams both with $b$ boxes, 
where $b\rightarrow \infty$ in the 't Hooft limit. This was based on a character analysis and used
the description of the coset representations in terms of plane partition. In this subsection, we 
approach this problem from a `microscopic' viewpoint, by studying the null-vector structure
of the relevant family of representations. As we shall see, this will nicely confirm the above results.
We shall concentrate on the ground state representations, i.e.\  the representations of the form
(\ref{bostwist1}) with $\ell=0$, since for them the analysis is simplest; the excitation spectrum will be studied in
more detail in the following subsection (albeit from a slightly different viewpoint).

\medskip

Since our considerations are only valid in the 't~Hooft limit we can decouple the outside-the-wedge modes
and think of these representations as representations of the wedge algebra ${\rm hs}[1]$. We are therefore
looking for the 
${\rm hs}[1]$ representation, whose character is the wedge character 
 $\chi_{[\nu]}^{(\ell)}$ with $\ell=0$, where
 \begin{equation}
\begin{aligned}
\chi_{[\nu]}^{(0)} (q) & = \left. q^h \prod_{n=1}^\infty (1-y\, q^{n-1+\nu})^{-1} (1-y^{-1} q^{n-\nu})^{-1}  \right|_{y^0} 
\\
& = q^h \, \left(1 + q + 3 q^2 + 6 q^3 + 12 q^4 + 21 q^5 + 38 q^6 + 63 q^7   + \cdots \right).\label{tw}
\end{aligned}
\end{equation}
Although the wedge character $\chi_{[\nu]}^{(0)}$ was computed using the free boson viewpoint  which
corresponds to hs$[1]$, we shall see below that  
an hs$[\lambda]$ representation with this character 
can be constructed for any value of $\lambda$, not just for $\lambda=1$.
One way to do this is to start with an arbitrary highest weight state $\phi$,
and determine the structure of the null-vectors it must possess in order to lead to a character 
of the form (\ref{tw}).

Let us denote the modes of the hs$[\lambda]$ algebra by $V^{(s)}_m$, where $s=2,3,\dots$ and $|m|\leq s-1$.
Furthermore, we denote by $w^{(s)}$ the eigenvalue of $V^{(s)}_0$ acting on a highest weight state $\phi$. 
A generic highest weight representation of  hs$[\lambda]$ can be specified by its charges $w^{(s)}$  
for all $s\geq 2$.  However, the representation given by  (\ref{tw}) is very special: it has only a single 
descendant at level one --- this type of representations was named `level-one representation' in \cite{Gaberdiel:2015wpo}.  
For a level-one representation, the condition that it has only a single state at level $1$ is so strong that it 
fixes all $w^{(s)}$ with $s\geq 4$ in terms of its conformal dimension $w^{(2)} \equiv h$ and its spin-$3$ 
charge $w^{(3)}$.

In order to see this, we first note that having only a single state at level $1$ means that all $V^{(s)}_{-1} \phi$ 
must be proportional to one another, and in particular, to $V^{(2)}_{-1}\phi$: 
\begin{align}\label{def-rel}
V^{(s)}_{-1}\phi = \frac{s w^{(s)}}{2h} V^{(2)}_{-1}\phi \ ,
\end{align}
where the proportionality factor is fixed by the requirement that these relations 
hold upon applying $V^{(2)}_1\equiv L_1$ to both sides.
Then, by taking the commutator with $V^{(3)}_0$, we can recursively determine the various
$w^{(s)}$ eigenvalues in terms of $h$ and $w^{(3)}$. It is more convenient to use $h$ and the ratio 
\be
\alpha \equiv \frac{3 w^{(3)}}{2 h} 
\ee 
to express the result, and for the first few spins we find explicitly 
\begin{equation}
\begin{aligned}
w^{(3)}&=\frac{2  h}{3}\alpha \label{alp-def}\\
w^{(4)}&=\frac{h}{2} \left[\alpha ^2+\frac{4-\lambda ^2}{20} \right]  \\
w^{(5)}&=\frac{2h}{5} \alpha\left[ \alpha^2 + \frac{20-3\lambda^2}{28} \right] \\
w^{(6)}&=\frac{h}{3} \left[	\alpha ^4+ \alpha ^2 \left(\frac{10-\lambda ^2}{6}\right) +\frac{\lambda ^4-20 \lambda ^2+64}{336} 	\right] \ .
\end{aligned}
\end{equation}
The expressions agree 
with those for the free boson obtained in eqs.~(B.5) -- (B.7) of \cite{Gaberdiel:2015wpo} upon setting $\lambda=1$.\footnote{Here 
we have worked with a different normalization of higher spin charges $w^{(s)}$ relative to that of 
\cite{Gaberdiel:2015wpo}: $w^{(s)}_{\textrm{here}}=4^{2-s} w^{(s)}_{\textrm{there}}$, and in particular, $\alpha_{\textrm{here}}=\frac{1}{4}\alpha_{\textrm{there}}$.}

\medskip
The level-one condition not only fixes all higher charges in terms of $(h,\alpha)$, together 
with the structure of hs$[\lambda]$ it also imposes very strong constraints on the number of descendants 
for every level. It has been shown recursively in \cite{Gaberdiel:2015wpo} that the wedge character of 
a generic level-one representation is precisely the MacMahon function (\ref{MacMahon}). 
More specifically, the full representation is generated by the modes $V^{(s)}_{-n}\phi$, where $s=n+1,\ldots,2n$,
and this matches then with 
another form of the MacMahon function, $M(q)= \prod^{\infty}_{n=1}\frac{1}{(1-q^n)^n} $.

Comparing now the $q$ expansion of the wedge character (\ref{tw}) with the MacMahon function (i.e.\
the hs$[\lambda]$ character of a generic level-one representation), we see that the ground state of the
twisted sector does not lead to a generic level-one representation: it has a first additional null-vector at level $4$ --- this was 
already noted in \cite{Gaberdiel:2015wpo}. This property can now be used to determine constraints on the 
parameters $(h,\alpha)$.
To do this, we study the structure of the null vectors systematically, from level $2$ up to level $5$.
Here we only give a brief summary of the results; the details can be found in 
Appendix~\ref{app:level1}. 

At each level, we have worked out the inner product matrix of the corresponding basis states and determined
its determinant. If this vanishes, this signals the appearance of a null-vector at that level. Any null-vector
at level $l$ will also give rise to descendant null-vectors at higher level; it is therefore of primary interest
to describe the new null-vectors that appear at each level. Up to level $5$, these new null-vectors arise
for the following values of $(h,\alpha)$:
\begin{equation}
\begin{aligned}\label{NRlevel23}
\hbox{Level 2:} \qquad& 
\alpha= \pm \Bigl( 1 \pm \frac{\lm}{2} \Bigr) \ , \quad \alpha = \pm \frac{1}{2} \sqrt{8h+\lm^2} \ ; \\
\hbox{Level 3:} \qquad &\alpha=\pm \Bigl( 2\pm \frac{\lm}{2} \Bigr)\ ,\quad  \alpha=\pm   \frac{1}{2} \sqrt{4h+\lm^2} \ ; \\
\hbox{Level 4:} \qquad
&\alpha=\pm \Bigl( 3\pm \frac{\lm}{2}\Bigr) \  ,\quad \alpha= \pm \frac{1}{2}{\sqrt{\tfrac{8}{3} h+ \lambda ^2}}\ ,		\\
&\alpha=\pm \frac{1}{2} \sqrt{\lambda^2-8h}\ ; \\
\hbox{Level 5:} \qquad& 
\alpha= \pm \Bigl( 4 \pm \frac{\lm}{2} \Bigr) \ , \quad \alpha = \pm \frac{1}{2} \sqrt{2h+\lm^2} \ .
\end{aligned}
\end{equation}
For the case at hand, we are interested  in the three new roots (not counting multiplicities and conjugations) at 
level $4$; as we have explained before, the ground state of the twisted sector has to satisfy (at least) one of these
relations.

All the roots, except for the third pair of roots at level $4$, follow a simple pattern, and as is explained in 
Appendix~\ref{app:level1},
all of these `standard' roots are attained by finite tensor powers of the minimal representation. It is thus very
suggestive that the additional null-vector that appears in the twisted sector representation corresponds to
this `special' root.

To confirm this, we have computed the value of $(h, \alpha)$ for the representation of the form
\be\label{bostwis0}
(\Lambda_+;\Lambda_-) = \bigl( [0^{b-1},1,0,\ldots,0]; [0^{b-1},1,0,\ldots,0] \bigr) \ ,
\ee
using the Drinfeld-Sokolov approach, see Appendix~\ref{app:B} for details. In particular, 
it follows from eqs.~(\ref{h-b}) and (\ref{alpha-b}), that the eigenvalues take the form 
 \begin{equation}\label{halphab}
\begin{aligned}
h 
&= \frac{\lambda^2 \, b (N+1) (N-b)}{2 N^2 (N+\lambda)}
\stackrel{\text{ `t Hooft}}{\approx} \ 
\frac{\lambda^2}{2} \Bigl(1-\frac{b}{N}\Bigr)\, \frac{b}{N}\\
\alpha 
&
= \frac{\lambda\,(N+2)(N -2b) }{2 N \sqrt{N(N+\lambda)}}  
\ \stackrel{\text{ `t Hooft}}{\approx} \ 
\frac{\lambda}{2} \Bigl(1-2\frac{b}{N}\Bigr)\ , 
\end{aligned}
\end{equation}
where we have first replaced $k$ in terms of $\lambda$, using (\ref{tHooft}), and then considered
the 't~Hooft limit.
Using  $0\leq \lambda \leq 1$ and demanding $0 <b < N$ --- only Young diagrams of  height at most $N$
are allowed --- we see that neither of the first two roots can be solved by representations of this type, whereas 
the last one 
\be\label{alpha} 
\alpha = \pm \frac{1}{2} \, \sqrt{\lambda^2 - 8 h} \ . 
\ee
is solved for any  $0< b < N$. If we further demand that $h>0$ in the 't Hooft limit (i.e.\  that they are not light states) 
then we conclude that\footnote{Note that under $\nu\rightarrow 1-\nu$, $(h,\alpha)\rightarrow (h,-\alpha)$ in 
eq.~(\ref{halphab}). Thus we may restrict ourselves to the range $0 < \nu <\frac{1}{2}$ if we include also
the conjugate representations.}
\begin{equation}\label{bofN}
b=\nu N \qquad \textrm{with}\qquad \nu <1 \ . 
\end{equation}
It only remains to understand the meaning of $\nu$, for which we go back to the special case of $\lambda=1$ 
(i.e.\  the free boson) from where we started. At $\lambda=1$, (\ref{halphab}) with (\ref{bofN}) reduces to
\begin{equation}
h=\frac{1}{2} \nu (1-\nu) \qquad \textrm{and}\qquad \alpha = \frac{1}{2}\, (1-2\nu)  \ ,
\end{equation}
which agrees with the result for the twisted sector representation
of the free boson from \cite{Gaberdiel:2010pz}, see its eq.~(4.4) and (4.5).\footnote{As remarked earlier in footnote 2, the normalization of 
$W^{(3)}$ differs by a factor of $4$ from that of \cite{Gaberdiel:2010pz}.} Thus 
we conclude that $\nu$ can indeed be identified with the twist parameter.

\subsection{The ground state conformal dimension and the excitation spectrum}\label{sec:exc}

In this section we confirm the identification between  the coset representations and the twisted sector states given 
in eq.~(\ref{twisgen}) (for the ground state) and (\ref{twistexc}) (for the excited states) by matching their conformal 
dimensions in the large $N$ limit. The analysis follows the same strategy as what was done for the 
${\cal N}=2$ case in \cite{Gaberdiel:2014vca}.
\smallskip

The conformal dimension of the ground state of the coset representation $(\Lambda_+,\Lambda_-)$ is 
\begin{equation}\label{h}
h(\Lambda_+;\Lambda_-) = \frac{C_2(\Lambda_+)}{N+k} +  \frac{C_2(\mu)}{N+1} -  \frac{C_2(\Lambda_-)}{N+k+1} +n \ , 
\end{equation}
where $C_2$ is the quadratic Casimir, $\mu$ is the $\mathfrak{su}(N)_1$ weight that is uniquely determined by the condition that 
$\Lambda_++  \mu - \Lambda_-$ lies in the root lattice, and $n$ denotes the first descendant level where 
$\Lambda_-$ appears in the affine representation of the numerator.
 The Casimir can be written in terms of the number of boxes in rows $r_i$ and columns $c_j$
as 
\begin{equation}\label{col-formula}
C_2(\Lambda) = \frac{1}{2} BN + \frac{1}{2} \left( \sum_{i} r_i^2 - \sum_j c_j^2   \right) - \frac{B^2}{2N} \ , 
\end{equation}
where $B=\sum_i r_i = \sum_j c_j$ is the total number of boxes.
\smallskip

Let us start with analysing the conformal dimension of the ground state of the twisted sector. In this case, 
since $\Lambda_+=\Lambda_-=\Lambda$, the $\mathfrak{su}(N)_1$ representation $\mu$ is trivial and $n=0$, 
and therefore the ground state conformal dimension equals
\be\label{hform}
h = \frac{C_2(\Lambda)}{(N+k) (N+k+1)}  \ . 
\ee
We are interested in the large $N$
behaviour at fixed $k$; since $r_i\leq k$, the $r_i^2$ term is subleading at large $N$, and since
the total number of boxes scales linearly with $N$, the same is true for the $B^2 / 2N$ term. Thus, to 
leading order in the $N\rightarrow \infty$ limit, we have 
\begin{align}\label{C2}
C_2 &\cong \frac{1}{2} BN - \frac{1}{2}   \sum_j c_j^2   
= \frac{N^2}{2} \sum_j \nu_j (1-\nu_j ) \ . 
\end{align}
Dividing by the denominator in (\ref{hform}) then leads to 
\begin{align}\label{bostwisg}
h  = \frac{1}{2} \sum_j \nu_j (1-\nu_j ) \ , 
\end{align}
which agrees precisely  with the  usual ground state energy of a  multi-twist sector
associated to the twist $(\nu_1,\ldots,\nu_k)$ with $\nu_1\geq \nu_2\geq \cdots \geq \nu_k$.

Having checked that the large $N$ limit of the coset representation $(\Lambda,\Lambda)$ of the form (\ref{twisgen}) 
agrees with the ground state energy of the twisted sector, we now compute the excitation spectrum above this ground 
state. For a generic twisted sector corresponding to $\nu_1\geq \nu_2\geq \cdots \geq \nu_k$, the different bosonic
generators will have mode numbers $-\nu_j + m$, where $j=1,\ldots, k$ and 
$m\in\mathbb{Z}$. Thus the lowest excitations raise
the conformal dimension by $\nu_j$. These different excitations should now correspond to the different ways
in which we can add a single box to $\Lambda_-$, without modifying $\Lambda_+$. 

To compute the difference in conformal dimension of the coset representation 
$(\Lambda;\Lambda^{(i)})$, where $\Lambda^{(i)}$ differs from $\Lambda$ by adding a single box to the 
$i$'th column, and that of the ground state $(\Lambda; \Lambda)$, we now use eq.~(\ref{h}). 
For $(\Lambda;\Lambda^{(i)})$, the $\mathfrak{su}(N)_1$ representation $\mu$ equals
the fundamental representation while $n$ remains $n=0$. Thus the difference in conformal dimension equals
\begin{align}\label{b-excite}
h(\Lambda;\Lambda^{(i)}) - h(\Lambda;\Lambda) & 
= \frac{C_2(\tyng(1))}{N+1} -  \frac{C_2(\Lambda^{(i)})-   C_2(\Lambda)}{N+k+1} \ . 
\end{align}
The Casimir of the fundamental is $C_2(\tyng(1)) = \frac{N^2-1}{2N}$, and the difference of the two Casimirs, 
to leading order in $N$, can be computed using  (\ref{C2})
\begin{align} \label{C2diff}
C_2(\Lambda^{(i)})-   C_2(\Lambda) & 
 \cong   \frac{N}{2} - c_i \ ,
\end{align}
where $c_i =  \nu_i N$ is the number of boxes in the $i$'th column of $\Lambda$.
Hence, in the $N \to \infty$ limit 
the excitation energy above the ground state takes the form
\begin{align}\label{bostwisnu}
h(\Lambda;\Lambda^{(i)}) - h(\Lambda;\Lambda) \cong \nu_i 
\end{align}
as expected. 

We should also note that if we remove a box from the $i$'th column of $\Lambda_{-}$, then the 
whole contribution from (\ref{C2diff}) changes sign, and the excitation energy is \linebreak
$\delta h = (1 - \nu_i)$.  This describes the action of the complex conjugate mode. These
results thus reproduce the excitation spectrum in the twisted sector, see, in particular,
eq.~(\ref{deltah}), where $\ell>0$  and $\ell <0$ corresponds to the excitation by the boson and its
complex conjugate, respectively. In terms of plane partitions, $\ell$ is the difference in height
of the window sills of $\Lambda_-$ relative to $\Lambda_+$; thus the action of the boson
and its conjugate can be thought of as adding a box to $\Lambda_-$ and $\Lambda_+$, respectively.\footnote{Since
the window sills have heights that are proportional to $N$, removing a box from $\Lambda_{-}$ is
equivalent to adding a box to $\Lambda_+$ in the large $N$ limit.} This identification is, however, only
valid in the large $N$ limit, where we have a `Fermi sea' of boxes and where the action of 
the anti-boxes can be described as creating a hole. In general, the plane partition viewpoint only describes
the representations that are made from boxes, and anti-boxes do not appear directly in this language.

\section{The twisted sector in the free fermion description}
\label{sec:3}

The bosonic coset theories also have a free field description for $\lambda=0$, where free fermions emerge. 
More precisely, $\lambda=0$ corresponds to taking $k\rightarrow \infty$ at fixed $N$, see eq.~(\ref{tHooft}), 
and the resulting theory can be identified with the ${\rm u}(1)$ coset of the theory of $N$ complex fermions, 
see \cite{Gaberdiel:2013jpa}. These fermions give rise to bilinear currents of the form 
\be
W^s(z) = n(s) \sum_{l=0}^{s-1} (-1)^l {s-1 \choose l}^2 \partial^{s-1-l} \psi^{\ast j} \,  \partial^{l} \psi^j \ , 
\ee
where $n(s)$ is an $s$-dependent normalization constant. The modes of these fields generate the 
linear ${\cal W}_{1+\infty}$ algebra \cite{Bergshoeff:1989ns,Bergshoeff:1990yd,Depireux:1990df}.
\smallskip

The bilinear currents are invariant under a ${\rm SU}(N)$ subgroup, and one can therefore think
of the resulting theory as a continuous orbifold by this group \cite{Gaberdiel:2011aa}. There are therefore again
twisted sectors that should admit a coset description. In the following we shall work out the details
of this correspondence. 
\smallskip

For the case of a single twisted complex fermion with modes
\begin{equation}
\psi_{r-\nu} \qquad \text{and}\qquad \psi^{\ast}_{r+\nu} \qquad \qquad \textrm{with} \quad r\in \mathbb{Z}+ \tfrac{1}{2} \ ,
\end{equation}
the wedge character is given by the fermionic analogue
of (\ref{twgen}),
\begin{align}\label{tw-f}
\phi_{[\nu]} (q,y) = q^h \prod_{n=1}^\infty (1+y\, q^{n-1/2+\nu}) (1+y^{-1}\, q^{n-1/2-\nu}) \ ,
\end{align}
where the twist is taken to lie in the interval $-\frac{1}{2} < \nu \leq \frac{1}{2}$. 
As for the bosonic case, the power of $y$ keeps track of the action under the cyclic group, 
and thus the states with a given power of $y$ furnish a representation of the hs$[0]$ algebra, with character
\begin{equation}
\phi^{(m)}_{[\nu]}(q) \equiv   \phi_{[\nu]} (q,y)\big|_{y^m}\ . 
\end{equation}
In contrast to the bosonic case, the fermionic wedge character $\phi^{(\ell)}(q)$ is much easier to compute. 
Using the Jacobi triple product identity, 
\begin{align}
\prod_{n=1}^\infty	(1-q^n)(1+y\, q^{n-1/2}) (1+y^{-1}\,q^{n-1/2}) = \sum_{n=-\infty}^{\infty} y^{n} q^{n^2/2} \ ,
\end{align}
we immediately have 
\begin{equation}\label{fermionwedge}
\begin{aligned}
\phi^{(m)}_{[\nu]}(q) 
&= q^{h+m \nu + m^2/2}\, \prod_{n=1}^\infty \frac{1}{(1-q^n) } \\
&= q^{h+m \nu + m^2/2} \bigl(1 +  q + 2 q^2 + 3 q^3 + 5 q^4 + 7 q^5 + 11 q^6+15 q^7 +\dots \bigr) \ .  
\end{aligned}
\end{equation}
In the following we shall use the plane partition viewpoint to identify the corresponding coset representations.

\subsection{The plane partition viewpoint}

Recall from the bosonic analysis of Section~\ref{sec:char1} that, from the plane partition perspective, the coset character
factorizes into two pieces: one corresponding to stacking boxes on the window sill --- this accounts precisely for the 
wedge character ---  and one corresponding to staking boxes on the empty floor --- this gives the 
MacMahon function that counts the outside-the-wedge modes. In the present situation we expect to find a similar
situation, except that now the wedge character $\phi^{(m)}(q)$ is the generating function of the partition number, i.e.\
it counts the number of Young diagrams with $n$ boxes. (This is in fact true for all $m$ --- the only $m$-dependence
of (\ref{fermionwedge}) appears in the overall exponent of $q$.)

The most naive guess for the correct plane partition configuration seems to be a single layer window sill
along the $x$-direction, say. In this scenario, stacking boxes on top of this window sill, which is equivalent 
to drawing usual two-dimensional Young diagrams, gives rise to the wedge character (\ref{fermionwedge}); 
whereas stacking boxes on the floor gives the MacMahon function, counting the outside-the-wedge modes of 
$\mathcal{W}_{1+\infty}$.
However, this guess turns out to be wrong since the coset representation corresponding to this plane partition, i.e.\
$([0^{b-1},1,0,\ldots,0];0)$ with $b\rightarrow \infty$, has conformal dimension (see eq.~(\ref{hsinglewall}))
\begin{equation}
h([0^{b-1},1,0,\ldots,0];0) = \frac{b (N-b)}{2N} \, \cdot \frac{2 N+k+1}{N+k}  \ , 
\end{equation}
which diverges in the 't~Hooft limit.\footnote{Naively, the only exception is the choice $b= N -a$, where $a$ is a finite positive integer, but this
then describes finitely many anti-boxes, i.e.\  does not have the correct coset character.}  

\begin{figure}[!h]
	\centering
	\begin{tabular}{c}
	         \includegraphics[width=.3\textwidth]{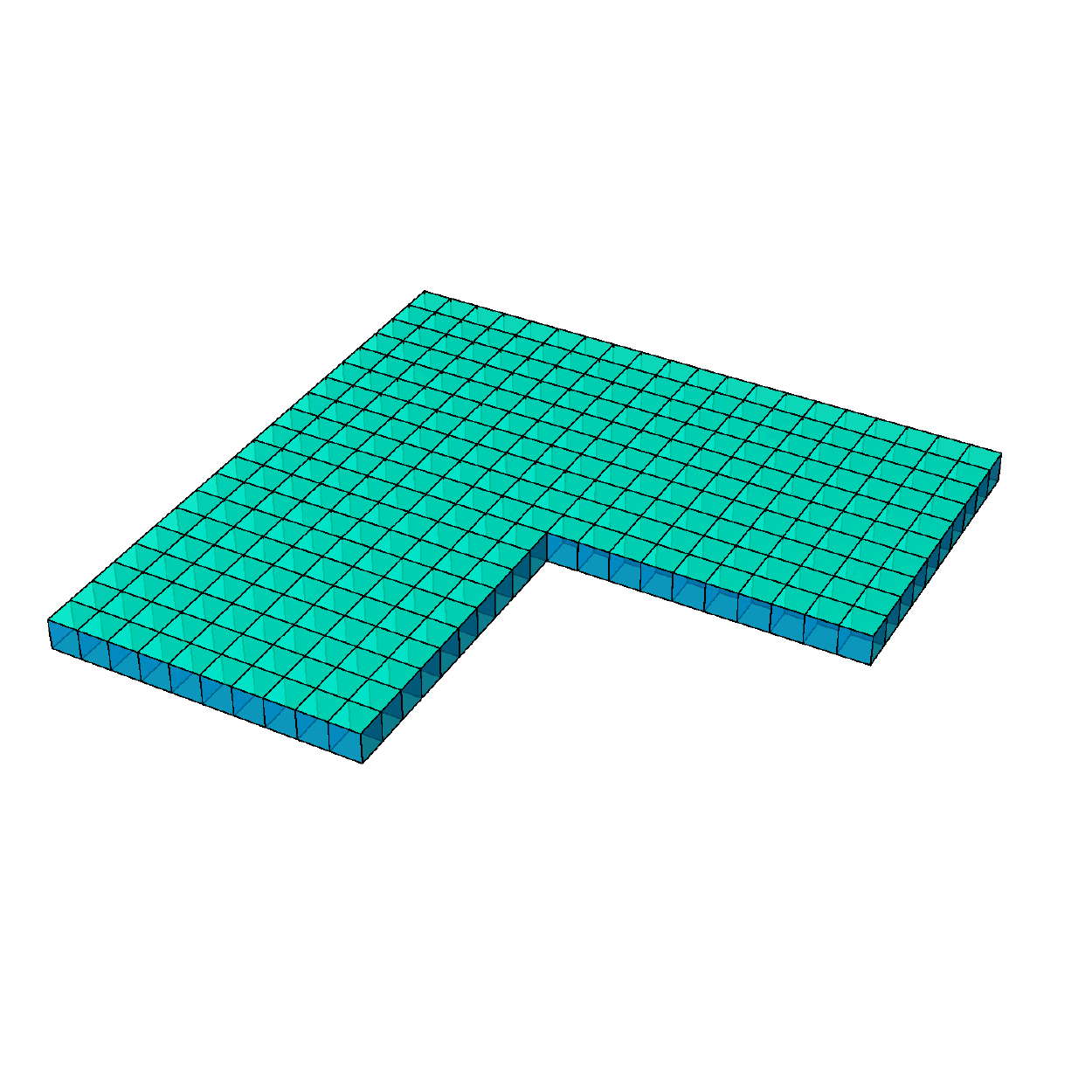} \qquad \qquad 
		\includegraphics[width=.3\textwidth]{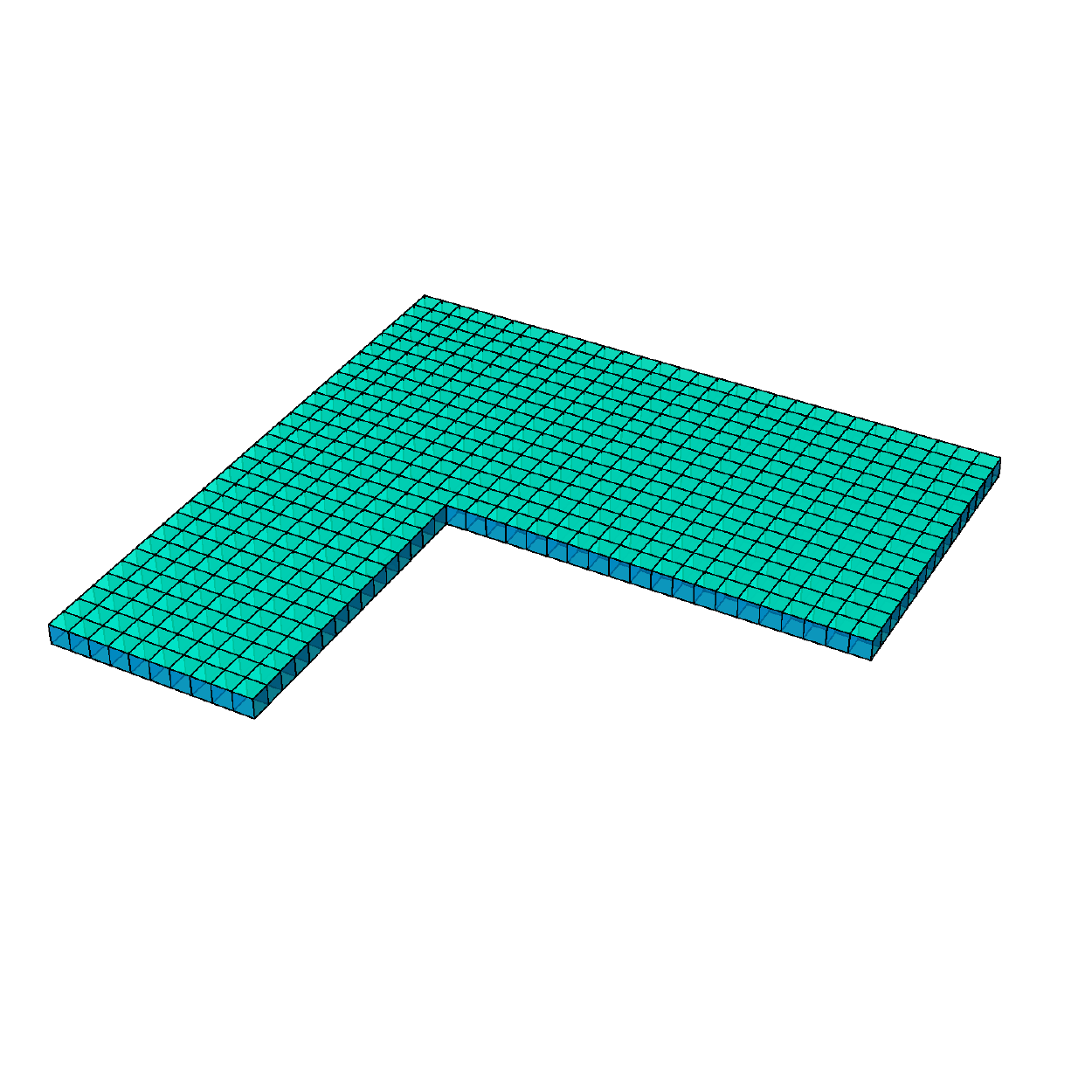}  \\ 
		\vspace{-1.5cm} (a) \hspace{5.5cm} (b) \vspace{1.5cm}
	\end{tabular}
	\caption{The  minimal configuration of plane partitions that describes the single-twist case for complex fermions. 
	Case (a) corresponds to the ground state representation, while (b) corresponds to the representation appearing at some excited level.}
	\label{fig3}
\end{figure}

However, we can also reverse the roles of the window sill and the floor, i.e.\  we can let the floor count the wedge character, 
and the window sill the outside-the-wedge modes. Then if we take the window sill to be only one box high, it can restrict the 
box-stacking on the floor to be a counting of Young diagrams. This leads us to the choice of plane partitions with a window-sill that
is only one box high, but has a fat `L' shape, whose widths $b$ and $b+m$ are taken to be large, see Figure~\ref{fig3}. In particular,
since the widths are large, the boxes that are placed on top of the window-sill, near the origin, are again counted by the 
MacMahon function (and hence describe the outside-the-wedge modes of  ${\cal W}_{1+\infty}$). On the other hand,
the configurations involving only boxes on the floor are counted again by plane partitions that satisfy a (generalized)
`pit' condition, where the pit is now located at $(x,y,z)=(1,1,2)$ and prevents any boxes from being stacked vertically.

The coset representations corresponding to this type of plane partition asymptotics are
\begin{align}
\bigl([b,0,\ldots,0];[b+ m, 0, \ldots,0] \bigr) 
\end{align}
for sufficiently large $b$. The neutral sector, whose wedge character is $\phi^{(0)}_{[\nu]}(q)$, corresponds to 
$m=0$, see Figure~\ref{fig3}\;(a); the higher representations with $m>0$ are shown in Figure~\ref{fig3}\;(b).

In all of these cases, the relevant representations are level-one representations that have one null-vector at level $2$,
i.e.\  there are only $2$ rather than $3$ states at the second descendant level. In fact, all of these representations
have $\alpha= \frac{1}{2} \sqrt{8h + \lambda^2}$ in the 't~Hooft limit, as follows from the analysis of appendix~\ref{app:B},
see e.g. eq.~(\ref{B.21}).

\begin{figure}[!h]
	\centering
	\begin{tabular}{c}
		\includegraphics[width=.4\textwidth]{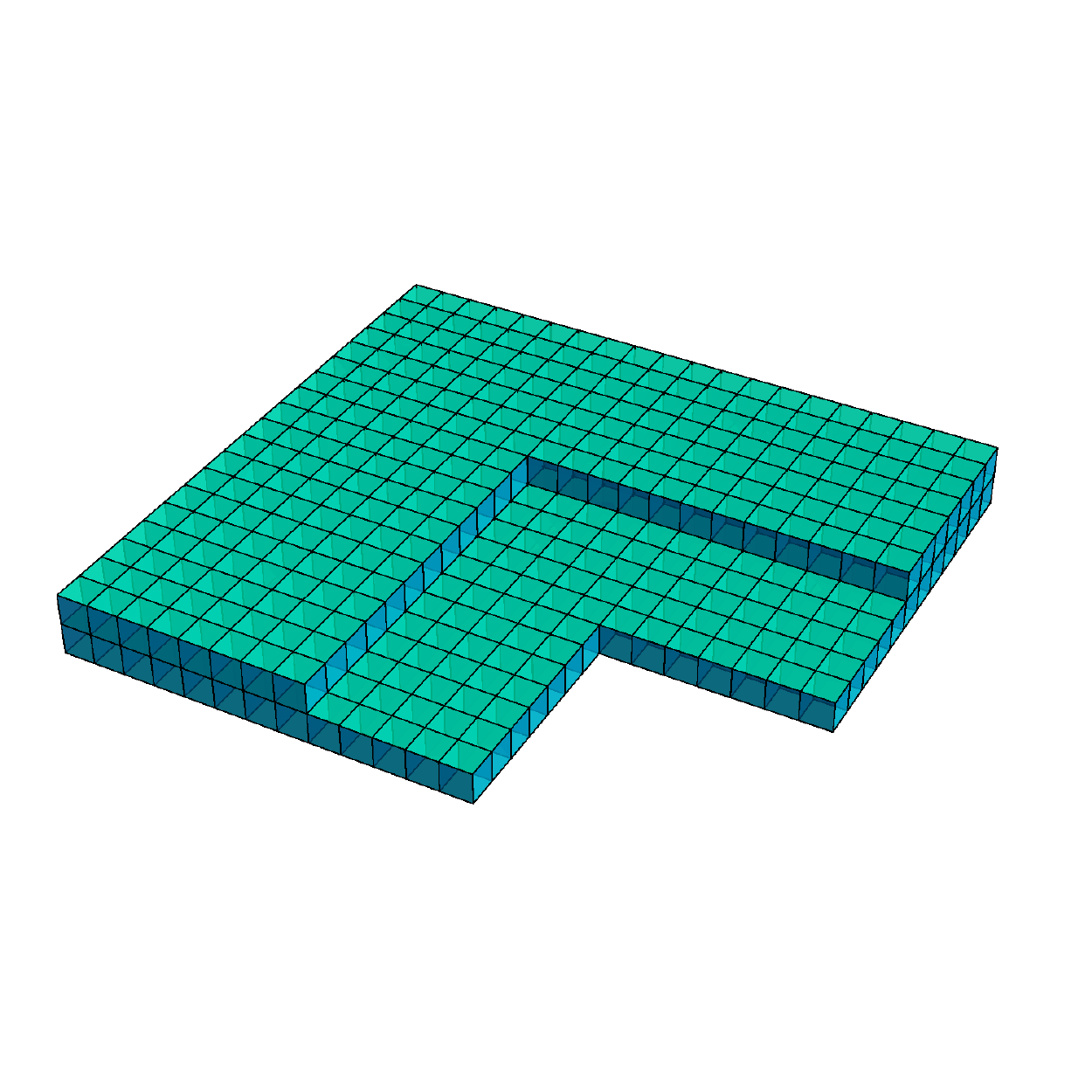} 
	\end{tabular}
	\caption{The  minimal configuration of plane partitions that corresponds to the ground state of the twisted sector where 
	two complex fermions are twisted.}
	\label{fig4}
\end{figure}

It is also not difficult to guess which plane partition asymptotics describe now the case where more than
one fermion is twisted: as in the bosonic case, we simply put different such diagrams together, see 
Figure~\ref{fig4} for the case where two fermions are twisted. 

The coset representation describing the ground state of a generic twist is then of the form 
\begin{align}\label{fertwisgen}
\bigl([b_1,b_2,b_3,\ldots,b_{N-1}];[b_1, b_2,b_3, \ldots,b_{N-1}]\bigr) \ , 
\end{align}
where the associated twists are 
\be\label{fertwist}
k \, \nu_i = \frac{B}{N} - r_i\ , 
\ee
and $r_i$ is the number of boxes in the $i$'th row, $i=1,\ldots, N$, while $B=\sum_i r_i$ is the total number of boxes.
This identification implies that the sum of all $N$ twists is zero, as has to be the case for a group element in ${\rm SU}(N)$.
We also note that by applying a field identification transformation, if necessary, we may assume that all $r_i\leq \frac{k}{2}$,
thus implying that $-\frac{1}{2} < \nu_1  \leq \nu_2 \cdots \leq \nu_N \leq \frac{1}{2}$.

\subsection{The ground state conformal dimension and the excitation spectrum}

As in the bosonic case, we can confirm these claims by direct CFT calculations. We begin with showing
that the coset representation (\ref{fertwisgen}) has the correct conformal dimension, namely
\be\label{fertwish}
h = \frac{1}{2} \sum_{i=1}^{N}  \nu_i^2 
\ee
in the $k\rightarrow \infty$ limit. Note that there are $N$ complex fermions, and each $\nu$-twisted 
fermion contributes $\nu^2/2$ to the ground state conformal dimension. In relating the free fermion theory to the coset
at $\lambda=0$, we need to take the $\mathfrak{u}(1)$ orbifold, i.e.\  we need to subtract from 
the conformal dimension of the numerator $-u^2/2N$, where $u$ is the $\mathfrak{u}(1)$-charge
of the state, see eq.~(\ref{u1dec}) below. Since $\sum_i \nu_i=0$, the $\mathfrak{u}(1)$-charge
is however $u=0$, and hence we arrive at (\ref{fertwish}). 

In order to derive this formula from the coset viewpoint, we note that the 
coset representation of the ground state is of the form $(\Lambda;\Lambda)$, 
where the row lengths  $r_i$ of $\Lambda$ satisfy (\ref{fertwist}). In the orthogonal basis of appendix~\ref{app:B},
see in particular eq.~(\ref{orthb}), the quadratic Casimir of $\Lambda$ is 
\begin{align}
C(\Lambda) & = \frac{1}{2}\, \sum_{i=1}^{N}  \Bigl(r_i -\frac{B}{N} \Bigr) \Bigl(r_i - \frac{B}{N} + N + 1 - 2i \Bigr) 
 \cong \frac{1}{2}\, \sum_{i=1}^{N}  \Bigl(r_i - \frac{B}{N} \Bigr)^2 = \frac{k^2}{2} \sum_{i=1}^{N} \nu_i^2   \ . 
\end{align}
Alternatively, this can also be seen from (\ref{col-formula}) since the $r_i^2$ and the $-B^2/2N$ terms  are the only 
expressions proportional to $k^2$. Dividing by $(N+k)(N+k+1) \cong k^2$ in the large $k$ limit, then leads to the 
desired expression for the conformal dimension, eq.~(\ref{fertwish}). 

Next, we study the excitation spectrum, following the same logic as above for the bosonic case, i.e.\  eq.~(\ref{b-excite}).  
From the difference of Casimirs we now find, to leading order in $k$, 
\begin{align}\label{Casfd}
C_2(\Lambda^{(i)})-   C_2(\Lambda) & \cong  \frac{1}{2}\, \bigl( (r_i+1)^2 - r_i^2 \bigr)  - \frac{(B+1)^2 - B^2}{2N}  \cong   r_i - \frac{B}{N} = - k \nu_i \ ,
\end{align}
where $\Lambda^{(i)}$ is now the representation that has an additional box in the $i$'th row. Thus the excitation spectrum
is 
\begin{align}\label{ferwisnu}
h(\Lambda;\Lambda^{(i)}) - h(\Lambda;\Lambda) \cong \nu_i + \frac{N-1}{2N} \ . 
\end{align}
For the free fermion theory one would have expected the answer to be $\nu_i + \frac{1}{2}$; the reason for the discrepancy
is that in order to obtain the ${\cal W}_\infty[0]$ theory that is described by the coset in the $k\rightarrow\infty$ limit, 
one has to divide out a $\mathfrak{u}(1)$ algebra --- this was explained in detail in \cite{Gaberdiel:2013jpa}. In particular,
the decoupled stress energy tensor is
\begin{align} \label{u1dec}
\tilde{T} = T - \frac{1}{2N} :JJ: \ , 
\end{align}
and since the individual fermions carry unit $\mathfrak{u}(1)$ charge, they define primary fields of conformal dimension
\begin{align}
\tilde{h}= \frac{1}{2}-\frac{1}{2N} = {N -1  \over 2N} 
\end{align}
in the decoupled theory. (Another way of reaching the same conclusion
is by observing that the conformal dimension of the $(0;\tyng(1))$ representation equals
\be
h(0;\tyng(1)) = \frac{(N-1)}{2N} \Bigl( 1 - \frac{N+1}{N+k+1} \Bigr) \cong \frac{(N-1)}{2N} 
\ee
in the large $k$ limit.) In either case, this then accounts precisely for the additional term in (\ref{ferwisnu}). 
Finally, we also note that removing a box from the $i$'th row changes simply the sign in (\ref{Casfd}), and 
thus leads to $-\nu_i + \frac{N-1}{2N}$ instead of (\ref{ferwisnu}). This then describes the action of the conjugate
fermionic mode, see eq.~(\ref{tw-f}).

\section{Conclusions}
\label{sec:4}

In this paper we have identified the twisted sector states of the bosonic higher spin CFTs in terms
of coset representations. The main idea 
was to use the description of ${\cal W}_\infty$ representations in terms of 
plane partitions \cite{FFJMM1,FFJMM2,FJMM1,Prochazka:2015deb}. Our analysis demonstrates that this method provides
a very powerful approach for the characterization of these representations, 
and this perspective is likely to have other important applications. 
\smallskip

The plane partition configurations that are crucial to this analysis  allow for a natural separation 
into the contribution associated to the wedge modes on the one hand, and those coming from the 
outside-the-wedge modes on the other, see e.g.\ the discussion in Section~\ref{sec:char1}. 
In particular, the former had a nice combinatorial description
in terms of plane partitions with a (generalized) pit condition. It would be interesting to see whether this observation
generalizes, and whether the representation theory of the wedge algebra ${\rm hs}[\lambda]$ can in general
be captured by plane partitions with suitable `pit' conditions.
\smallskip

At present, the plane partition viewpoint has only been developed for the ${\cal W}_\infty$
algebra that appears in the duality to bosonic higher spin theory; it would be very interesting
to generalize this technique to the supersymmetric cases. In particular, the ${\cal N}=2$ case
where Young super tableaux \cite{sergeev,berele} naturally appear (see e.g.\ \cite{Candu:2012jq})
should allow for a nice generalization. 

Given that the plane partitions
also describe the representation theory of the affine Yangian algebra of $\mathfrak{gl}_1$, which is 
believed to contain ${\cal W}_\infty$ as a subalgebra \cite{Prochazka:2015deb}, this viewpoint
relates higher spin symmetries to Yangian symmetries that typically arise  in integrable systems. This 
approach may therefore pave the way towards understanding the relation between higher spin symmetries
and integrability. As with the embedding into string theory, it is likely that the sharpest results 
will be possible in the maximally supersymmetric ${\cal N}=4$ case, and hence it would be very 
interesting to find the appropriate ${\cal N}=4$ supersymmetric generalization of the affine Yangian. 
\smallskip

The coset representations we have found exist for generic values of $N$ and $k$, while a direct twisted 
sector interpretation is only possible for the free field cases which correspond to $\lambda=0$ (free fermions) 
and $\lambda=1$ (free bosons). It would be interesting to understand whether these representations also
have a natural interpretation away from these points, for example in terms of parafermions. Finally, 
it is intriguing that the structure of the coset representations is very similar in both cases to 
those representations that appear in the extension from the higher spin algebra to the Higher Spin Square,
see  eqs.~(3.1) and (3.10) in \cite{Gaberdiel:2015mra}. It would be very interesting to understand
the reason underlying this.

\section*{Acknowledgement}

We thank Kevin Ferreira, Rajesh Gopakumar, Maximilian Kelm, Tomas Proch\'{a}zka, and Huafeng Zhang for helpful discussions. The research of SD is supported
by a grant from the NCCR SwissMAP, funded by the Swiss National Science Foundation. We thank  ICTP, Trieste and MIAPP, Munich for hospitality while part of this work was done. MRG and CP thank ITP of Chinese Academy of Science for hospitality, while WL thanks the ITP of ETH Zurich for hospitality.

\appendix

\section{The general form of level-1 representations}\label{app:level1}

In this appendix we study the structure of a generic level-$1$ representation at arbitrary $\lambda$. This
can be done level by level from their Kac determinants. In the following, $V^{(s)}_m$ denotes the 
hs$[\lm]$ generator of spin $s$ and mode $m$.

\subsection*{Level 2}

At level $2$, a general level-$1$ representation has three ${\rm hs}[\lambda]$ descendants, namely
\begin{align}
V^{(3)}_{-2} \phi \quad , \quad  V^{(4)}_{-2} \phi \quad , \quad {V^{(2)}_{-1}}^{ 2} \phi  \ . 
\end{align}
[It is not difficult to show that any other ${\rm hs}[\lambda]$ descendant can be written as 
a linear combination of these, using the commutation relations of the ${\rm hs}[\lambda]$
algebra, as well as the fact that there is only a single state at level-$1$.]
We have worked out the inner product matrix of these states, and its determinant is of the form 
\begin{align}
\det(M_2)= 16h^3 \left(\alpha-1-\tfrac{\lm}{2}  \right) \left(\alpha-1+\tfrac{\lm}{2}  \right) \left(\alpha+1-\tfrac{\lm}{2}  \right) 
 \left(\alpha+1+\tfrac{\lm}{2}  \right) \left(2 h - \alpha^2 + \tfrac{\lm^2}{4}   \right) \ . 
\end{align}
Thus the zeros appear at 
\begin{align}\label{level2zero}
\hbox{Level 2:} \qquad 
\alpha= \pm \Bigl( 1 \pm \frac{\lm}{2} \Bigr) \ , \quad \alpha = \pm \frac{1}{2} \sqrt{8h+\lm^2} \ , 
\end{align}
where each root has a single multiplicity (and for the first expression the two minus signs are uncorrelated, i.e.\
this describes 4 different roots). The overall sign of $\alpha$ relates conjugate representations to one another
(since the eigenvalue of $V^{(3)}_0$ has opposite sign for conjugate representations, while the conformal dimension
remains the same). 

These roots have a simple interpretation in terms of familiar coset representations. It follows from the 
analysis of \cite{Gaberdiel:2011wb} that the minimal representations $(\tyng(1);0)$ and $(0;\tyng(1))$ (as well
as their conjugates) have the eigenvalues
\begin{align}
&h\bigl((\tyng(1);0)\bigr) = \frac{1}{2} (1+\lambda) \ , \qquad   w^{(3)}\bigl((\tyng(1);0)\bigr) =  - \frac{1}{6} (1+\lambda) (2+\lambda) \ , \\
&h\bigl((0;\tyng(1))\bigr) = \frac{1}{2} (1-\lambda) \ , \quad \ \  \  w^{(3)}\bigl((0;\tyng(1))\bigr) =   \frac{1}{6} (1-\lambda) (2-\lambda) \ .
\end{align}
The corresponding $\alpha$ values are then 
\be
\alpha\bigl((\tyng(1);0)\bigr) = - \Bigl(1 + \frac{\lambda}{2} \Bigr) \ , \qquad
\alpha(\bigl(0;\tyng(1)) \bigr) = \Bigl( 1 - \frac{\lambda}{2} \Bigr) \ .
\ee
These therefore account for the first four zeros of (\ref{level2zero}), including the corresponding conjugate representations.
However, they actually also solve the last two equations of (\ref{level2zero}) since for $h = \frac{1}{2} (1\pm \lambda)$,
\be
\frac{1}{2} \sqrt{8h + \lm^2}  = \frac{1}{2} \sqrt{4 \pm 4\lm + \lm^2} = \pm \frac{1}{2}  (2\pm \lm) \ . 
\ee
This reflects the fact that the minimal representations $(\tyng(1);0)$ and $(0;\tyng(1))$ (or their conjugates) have {\em two}
null-vectors at level $2$ --- they only have a single wedge descendant at this level. 

In order to identify the representations that have only a single null-vector (corresponding to a single zero) at level $2$, we 
note that the symmetric tensor powers of the minimal ${\rm hs}[\lambda]$ representations\footnote{Recall that there 
is a transpose in relating the symmetrization or anti-symmetrization of the ${\rm hs}[\lambda]$ representations to the 
Young diagrams appearing the coset, see the discussion in section~2.2 of \cite{Gaberdiel:2011zw}, e.g. eq.~(2.18).}
satisfy 
\begin{align}
&h\bigl(([0^{m-1},1,0,\ldots,0];0)\bigr) = \frac{m}{2} (1+\lambda) \ , \nonumber  \\
&\qquad \qquad w^{(3)}\bigl(([0^{m-1},1,0,\ldots,0];0)\bigr) =  - \frac{m}{6} (1+\lambda) (2+\lambda) \ , \label{A.14}\\
&h\bigl((0;[0^{m-1},1,0,\ldots,0])\bigr) = \frac{m}{2} (1-\lambda) \ , \nonumber \\
& \qquad \qquad w^{(3)}\bigl((0;[0^{m-1},1,0,\ldots,0])\bigr) =   \frac{m}{6} (1-\lambda) (2-\lambda) \ . \label{A.15}
\end{align}
These representations therefore account for one of the first four zeros of (\ref{level2zero}), including also 
the corresponding conjugate representations. However, for $m\geq 2$, they do not satisfy the last two zeros any longer.
This is compatible with the fact that the corresponding wedge representations have two states at level $2$, i.e.\  only have a single
null-vector at that level.

On the other hand, a representation for which only one of the last two zeros is satisfied is given by the two-fold anti-symmetric
tensor power of the minimal representation. Indeed, it follows from appendix B.3 of \cite{Gaberdiel:2011zw} that for example
\be
h\bigl(([2,0,\ldots,0];0)\bigr) = 2 + \lambda \ , \qquad
w^{(3)} \bigl(([2,0,\ldots,0];0)\bigr) = -\frac{1}{3} (2+\lambda) (4+\lambda) \ , 
\ee
where we have noted that the normalisation of $W^{(3)}$ in \cite{Gaberdiel:2011zw} differs by a factor of $\frac{1}{6}$ from 
the conventions of \cite{Gaberdiel:2011wb}, see in particular eq.~(5.8) of \cite{Gaberdiel:2011zw}. Thus $\alpha$
takes the value
\be\label{anti2}
\alpha\bigl(([0,1,0,\ldots,0];0)\bigr) = - \frac{1}{2}\, (4 + \lambda)  = - \frac{1}{2} \, \sqrt{8 (2+\lambda) + \lambda^2} \ . 
\ee
 
\subsection*{Level 3}
 
We have similarly determined the Kac-determinant (and in particular its zeros) at higher levels. At level $3$, there are  
generically $6$ independent vectors, for which a basis is given by 
\begin{align}
\V{4}{-3}\phi \ \ , \ \ \V{5}{-3}\phi \ \ , \ \ \V{6}{-3}\phi \ \ , \ \  \V{2}{-1}\V{3}{-2}\phi \ \ , \ \  \V{2}{-1}\V{4}{-2}\phi \ \ , \ \  \V{2}{-1}\, \V{2}{-1} \,\V{2}{-1}\phi \ \ . 
\end{align}
The corresponding Kac determinant is then 
\begin{align}
\det(M_3) = &\frac{3 h^6}{16384} (2 \alpha -\lambda -4) (2 \alpha -\lambda -2)^3 (2 \alpha -\lambda +2)^3 
(2 \alpha -\lambda +4) (2 \alpha +\lambda -4) \nonumber \\ 
&  \qquad \quad (2 \alpha +\lambda -2)^3
(2 \alpha +\lambda +2)^3 (2 \alpha +\lambda +4) \left(-4 \alpha ^2+4 h+\lambda ^2\right)  \nonumber \\
& \qquad \quad \left(-4 \alpha ^2+8 h+\lambda ^2\right)^3 \ ,
\end{align}
and its $24$ zeros arise for 
\begin{align}
\alpha =  &  \pm \Bigl(1 \pm \frac{\lm}{2} \Bigr) \ , \quad  \pm  \frac{1}{2} \sqrt{8h+\lm^2} , \nonumber \\ 
&\pm \Bigl( 2\pm \frac{\lm}{2} \Bigr)\ ,\quad  \pm  \frac{1}{2} \sqrt{4h+\lm^2} \ .  
\end{align}
Here each of the zeros that appeared already at level $2$ --- these are the zeros of the first line --- has multiplicity $3$, 
while the new zeros have multiplicity one. The new roots are satisfied for the anti-symmetric two-fold tensor product
of the minimal representation, see e.g. eq.~(\ref{anti2}) --- together with the representation associated to the other minimal
representation as well as their conjugates, this accounts for the first four new roots. The last two new roots
are attained for the $2$-fold symmetric tensor power of the minimal representation, see eqs.~(\ref{A.14}) 
and (\ref{A.15})  with $m=2$.

\subsection*{Level 4}

At level $4$, there are generically $13$ states, and the $78$ roots of the corresponding Kac determinant (including multiplicities) are 
\begin{align}
\alpha =  &  \pm  \Bigl( 1 \pm \frac{\lm}{2} \Bigr) \  , \quad   \pm  \frac{1}{2} \sqrt{8h+\lm^2} \ , \label{41} \\
&\pm \Bigl( 2\pm \frac{\lm}{2} \Bigr)\ ,\quad  \pm  \frac{1}{2} \sqrt{\tfrac{8}{2} h+\lm^2}\ , \label{42}   \\
&\pm \Bigl( 3\pm \frac{\lm}{2}\Bigr) \  ,\quad  \pm \frac{1}{2}{\sqrt{\tfrac{8}{3} h+ \lambda ^2}}\ ,	\label{43}	\\
&\pm \frac{1}{2} \sqrt{\lambda^2-8h}\ . \label{44}
\end{align}
The roots in (\ref{41}) appear already at level $2$, while those in (\ref{41})  and (\ref{42}) appear at level $3$; these roots therefore
each have higher multiplicity.\footnote{The 4 roots $\pm ( 1 \pm \frac{\lm}{2})$ each have multiplicity $9$, while the 2 roots 
$\pm  \frac{1}{2} \sqrt{8h+\lm^2}$ have multiplicity $8$. The 6 roots in (\ref{42}) all have multiplicity $3$.}
 The new roots (that appear with multiplicity one) are therefore associated to the solutions in (\ref{43}) and (\ref{44}). 
The roots in (\ref{43}) are associated to the totally anti-symmetric three-fold tensor product of the minimal representation --- this accounts for the 
first four roots of (\ref{43}) --- and the totally symmetric three-fold tensor product of the minimal representation --- this accounts for the last 
two roots of (\ref{43}). On the other hand, the roots in line (\ref{44}) are of a different form, and they are the ones that are relevant for the bosonic 
twisted representation, see eqs.~(\ref{h-b}) -- (\ref{alpha-b}).

\subsection*{Level 5}

We have also performed the corresponding analysis at level $5$, where the generic level-one representation has $24$ states,
and where the Kac determinant has 192 roots (including multiplicities). In addition to the roots that appeared already at level $4$, 
see eqs.~(\ref{41}) -- (\ref{44}), the new roots that appear at level $5$ are of the form
\be
\pm \Bigl(4 \pm \frac{\lambda}{2} \Bigr) \ , \qquad \pm \frac{1}{2}  {\sqrt{\tfrac{8}{4} h+ \lambda ^2}}\ , 
\ee
and are hence of the same structural form as in (\ref{41}) -- (\ref{43}). In particular, they correspond to the 
totally symmetric and anti-symmetric four-fold tensor product of the minimal representation.

\section{The spin 3 charge of some simple representations}\label{app:B}

In order to identify the appropriate representations  that realize these roots, we need to calculate both the conformal
dimension as well as the $w^{(3)}$ eigenvalue of level-one representations. In this section we recall how the spin $3$
charge can be determined, using the Drinfeld-Sokolov approach.

The coset representations are labelled by pairs of $\mathfrak{su}(N)$ representations $(\Lambda_+;\Lambda_-)$. 
For each such $\Lambda$, we denote by $r_i$ the number of boxes in the $i$'th row of the corresponding Young diagram.
Then, in the orthonormal basis, the weight $\Lambda$ has components (see e.g. appendix A of \cite{Gaberdiel:2011zw}) 
\begin{align}
\Lambda_i = r _i - \frac{B}{N} \ , \qquad i = 1,\ldots, N \ , \label{orthb}
\end{align} 
where $B$ is the total number of boxes $\sum_i r_i$ and ($r_N \equiv 0$). The Weyl vector $\rho$ has the components 
\begin{align}
\rho_i = \frac{N+1}{2} - i\ ,  \qquad  i = 1,\ldots, N \ . 
\end{align}
Note that, by construction, we have $\sum_i \Lambda_i =0$ and $\sum_i \rho_i =0$. Following \cite{Perlmutter:2012ds}, we 
also define the vector  $\th$ as
$$
\th = \alpha_+ (\Lambda^+ + \rho) + \alpha_- (\Lambda^- +\rho) \ ,  \quad \text{where} \quad 
\alpha_+= \sqrt{\frac{N+k+1}{N+k}}\ ,  \quad \alpha_- =- \frac{1}{\alpha_+} \ ,
$$
and introduce the power sums $C_s(\th)$ \cite{Bouwknegt:1992wg,Castro:2011iw,Perlmutter:2012ds}
\begin{align}
C_s(\th) = \frac{1}{s}\sum_i (\th_i)^s \  \ = \frac{(-1)^{s-1}}{s} \sum_{i_1 <i_2 < \cdots < i_s} \th_{i_1}\th_{i_2}\cdots \th_{i_s} \ . 
\end{align}
In terms of these quantities, the conformal dimension and spin $3$ charge of $(\Lambda_+; \Lambda_-)$ in the `primary basis'
is then  
\cite{Bouwknegt:1992wg,Perlmutter:2012ds}
\begin{align}
h \bigl( (\Lambda_+; \Lambda_-) \bigr) & = C_2 (\th) + \frac{c-N+1}{24} \label{hDS} \\
w^{(3)} \bigl( (\Lambda_+; \Lambda_-) \bigr) & =  C_3(\th) \ . \label{wDS} 
\end{align}
As a consistency check we note that, for the minimal representation $\Lambda_+=\tyng(1)$, we find 
\begin{align}
h(\tyng(1);0) & = \frac{N-1}{2N} \, \frac{1+k+2 N}{N+k} \\
w^{(3)}(\tyng(1);0) & = \frac{(N-1)(N-2)}{6N^2} \, \frac{(1+k+2 N)(2 + 2k + 3N)}{(N+k)^{3/2} (N+k+1)^{1/2}} \ .
\end{align}
Similarly, we find for example for the representation $([0^{b-1},1,0,\ldots,0];0)$ 
\begin{align}\label{hsinglewall}
h([0^{b-1},1,0,\ldots,0];0)& = \frac{b (N-b)}{2N} \, \frac{1+k+2 N}{N+k} \\
w^{(3)}([0^{b-1},1,0,\ldots,0];0) & = \frac{b (N-b)(N-2b)}{6N^2} \, \frac{(1+k+2 N)(2 + 2k + 3N)}{(N+k)^{3/2} (N+k+1)^{1/2}} \ , 
\end{align}
so that in the 't~Hooft limit (for finite $b\ll N,k$) 
\begin{align}
h([0^{b-1},1,0,\ldots,0];0)& = b \, h(\tyng(1);0)  \cong  \frac{b}{2} (1+\lambda) \\
w^{(3)}([0^{b-1},1,0,\ldots,0];0) &= b \, w^{(3)}(\tyng(1);0) \cong \frac{b}{6} (1+\lambda) (2+\lambda) \ . 
\end{align}

This approach also allows us to calculate the charges of representations for which both $\Lambda_+$ and
$\Lambda_-$ are non-trivial. For example, for the representation $(\tyng(1);\tyng(1))$ we find 
\begin{align}
h(\tyng(1);\tyng(1)) &= \frac{(N-1)(N+1)}{2 N (N+k) (N+k+1)} \\
w^{(3)} (\tyng(1);\tyng(1)) &= \frac{(N-2) (N-1) (N+1) (N+2)}{6 N^2 (k+N)^{3/2} (k+N+1)^{3/2}} \\
\alpha(\tyng(1);\tyng(1)) &= \frac{3 w^{(3)}}{2h} = \frac{(N-2) (N+2)}{2 N \sqrt{k+N} \sqrt{k+N+1}} 
\stackrel{\text{ `t Hooft}}{\approx} \frac{1}{2} \frac{N}{N+k} = \frac{\lm}{2} \ , 
\end{align}
where we have taken the 't~Hooft limit in the last step. Note that the correct value of $\alpha$
in the 't~Hooft limit 
can only be determined from the exact expression of $h$ and $w^{(3)}$ at finite $(N,k)$ since
the 't~Hooft limit of both $h$ and $w^{(3)}$ separately vanishes. 

\smallskip
For the representations that describe the ground states of a single twisted boson, i.e.\
the representations with Dynkin labels $([0^{b-1},1,0,\ldots,0];[0^{b-1},1,0,\ldots,0])$,
we then find 
\begin{align}
h &=  \frac{b (N+1) (N-b)}{2 N (k+N) (k+N+1)} \label{h-b} \\
\alpha 
& = \frac{(N -2b) (N+2)}{2 N \sqrt{k+N} \sqrt{k+N+1}}  \ . \label{alpha-b}
\end{align}
As explained in eq.~(\ref{halphab}), in the 't~Hooft limit this solves the root of eq.~(\ref{44}) that 
appears first at level~$4$.
\medskip

For the representations that describe the ground states of a single twisted fermion, i.e.\
the representations with Dynkin labels  $([b,0,\ldots,0];[b,0,\ldots,0])$, we find instead
\begin{align}
h&=  \frac{b (N-1) (N+b)}{2 N (k+N) (k+N+1)} \\
\alpha &  
= \frac{(N +2b) (N-2)}{2 N \sqrt{k+N} \sqrt{k+N+1}}  \ . 
\end{align}
Upon setting $b=\nu k$, we have (in the `t\,Hooft limit) 
\begin{align}
h (b=\nu k) &  
\cong \frac{\nu}{2} (1-\lambda) ( \lambda (1-\nu) + \nu) \\
\alpha (b=\nu k)  & 
\cong \frac{1}{2} (\lambda(1 - 2\nu) + 2\nu) = \frac{1}{2}\,
 \sqrt{\lambda^2+8h (b=\nu k)} \ , \label{B.21}
\end{align}
which is a root that first appears at level $2$, see eq.~(\ref{level2zero}).

\section{Combinatorial description of wedge characters}
\label{app:C}

In this appendix we outline a proof for the combinatorial identity between $p_2(n,\ell)$ and $p_2(n)$ of
eq.~(\ref{p2rel}). The identity for $\ell=0$ is given and proven in  \cite{combinatorics}; in the following we
will generalize it to generic $\ell$. Let us first consider the case of $\ell\geq 0$. 

Recall that $p_2(n,\ell)$ counts the configurations that can be obtained from a pair of Young diagrams 
$\langle\Gamma^+;\Gamma^-\rangle_{n,\ell}$ with first columns of height $c_1^+ = c_1^- + \ell$ by gluing them 
along their common first column, and removing the $\ell$ superfluous boxes of $\Gamma^+$ from the bottom,
see Figure~\ref{overlap-young}. Here $n$ is the number of `visible' boxes after the gluing.

On the other hand, $p_2(n)$ counts (ordered) pairs of Young diagrams 
$(\Gamma^{(1)},\Gamma^{(2)})_n$ whose total number of boxes is $n$. To each such pair
we can associate an element $\langle\Gamma^+;\Gamma^-\rangle_{n,\ell}$, by shifting $\Gamma^{(2)}$ 
$\ell$ steps upwards and then placing it to the right of $\Gamma^{(1)}$, without letting any columns overlap. 
As long as $c_1^{(1)} \leq c_1^{(2)}+\ell$, we choose the first  column of $\Gamma^{(2)}$ as the `shared' column  of $\langle\Gamma^+;\Gamma^-\rangle_{n,\ell}$; otherwise we move $\ell$ boxes from the bottom of the first column of  $\Gamma^{(1)}$ to the top so that this column becomes the `shared' column. 

This map is well-defined and surjective, but 
it is not injective. In particular, if $c_1^{-} > c_2^{-}+\ell$ there are precisely two pairs of Young 
diagrams $(\Gamma^{(1)},\Gamma^{(2)})_n$ that give rise to the same configuration 
$\langle\Gamma^+;\Gamma^-\rangle_{n,\ell}$ --- we can move $\ell$ boxes from the top of the
first column of $\Gamma^{(2)}$ to the bottom, and then adjoin the corresponding column to 
$\Gamma^{(1)}$. Therefore, if we start with $p_2(n)$, we now have to subtract $p_2(n-\ell-1)$ from it, since
the diagrams that are overcounted in this manner  can all be constructed from configurations 
$(\Gamma^{(1)},\Gamma^{(2)})_{n-\ell-1}$:  we simply add $\ell+1$ boxes to the first column of $\Gamma^{(2)}$, and
thus guarantee that after joining we have $c_1^{-} > c_2^{-}+\ell$ in the resulting $\langle\Gamma^+;\Gamma^-\rangle_{n,\ell}$ 
configuration. But subtracting $p_2(n-\ell-1)$ is an overkill --- we now have to add back those configurations for which
$c_1^->c_2^{-}+\ell>c_3^{-}+\ell$, and these cases are captured by $(\Gamma^{(1)},\Gamma^{(2)})_{n-(\ell+1) - (\ell+2)}$ since 
now we have to put $\ell+1$ boxes on the second column of $\Gamma^{(2)}$, and $\ell+2$ boxes on the first so as to guarantee
that we end up with a configuration with $c_1>c_2^{-}+\ell>c_3^{-}+\ell$.
Recursively proceeding in this manner we then arrive at the formula 
\begin{equation}
p_2(n,\ell)=\sum_{m=0}(-1)^m\,p_2 \Bigl(n-\sum^{\ell+m}_{k=\ell+1}k \Bigr)  \qquad \qquad \ell\geq 0\ . 
\end{equation}
This proves eq.~(\ref{p2rel}) for $\ell\geq 0$; the argument for $\ell<0$ is identical 
upon interchanging the roles of $\Gamma^+ \leftrightarrow \Gamma^-$ and $\Gamma^{(1)} \leftrightarrow\Gamma^{(2)}$.

\bibliographystyle{JHEP}

\begin{thebibliography}{99}

\bibitem{Vasiliev:2003ev}
M.A.~Vasiliev,
``Nonlinear equations for symmetric massless higher spin fields in (A)dS(d),"
Phys.\ Lett.\  B {\bf 567} (2003) 139
{\tt [arXiv:hep-th/0304049]}.

\bibitem{Sundborg:2000wp}
B.~Sundborg,
``Stringy gravity, interacting tensionless strings and massless higher spins,"
Nucl.\ Phys.\ Proc.\ Suppl.\  {\bf 102} (2001) 113
{\tt [arXiv:hep-th/0103247]}.
  
\bibitem{Witten}
E.~Witten, talk at the John Schwarz 60-th birthday symposium (Nov. 2001), \newline
{\tt http://theory.caltech.edu/jhs60/witten/1.html}.

\bibitem{Mikhailov:2002bp}
A.~Mikhailov,
``Notes on higher spin symmetries,''
{\tt arXiv:hep-th/0201019}.


 \bibitem{Klebanov:2002ja}
 I.R.~Klebanov and A.M.~Polyakov,
``AdS dual of the critical O(N) vector model,"
Phys.\ Lett.\  B {\bf 550} (2002) 213
{\tt [arXiv:hep-th/0210114]}.


\bibitem{Sezgin:2003pt} 
E.~Sezgin and P.~Sundell,
``Holography in 4D (super) higher spin theories and a test via cubic scalar couplings,''
JHEP {\bf 0507} (2005) 044
{\tt  [arXiv:hep-th/0305040]}.
 

\bibitem{Gaberdiel:2010pz}
M.R.~Gaberdiel and R.~Gopakumar,
 ``An AdS$_3$ dual for minimal model CFTs,''
Phys.\ Rev.\ D {\bf 83} (2011) 066007
 {\tt [arXiv:1011.2986 [hep-th]]}.
 
\bibitem{Chang:2012kt} 
C.-M.~Chang, S.~Minwalla, T.~Sharma and X.~Yin,
``ABJ triality: from higher spin fields to strings,'' 
J.\ Phys.\ A: Math.\ Theor.\ {\bf 46} (2013) 214009 
{\tt [arXiv:1207.4485 [hep-th]]}.

 
\bibitem{Gaberdiel:2014cha} 
M.R.~Gaberdiel and R.~Gopakumar,
``Higher Spins \& Strings,''
JHEP {\bf 1411} (2014) 044 
{\tt [arXiv:1406.6103 [hep-th]]}.


\bibitem{Gaberdiel:2013vva}
M.R.~Gaberdiel and R.~Gopakumar,
``Large $\mathcal{N}=4$ holography,''
 JHEP {\bf 1309} (2013) 036
{\tt  [arXiv:1305.4181 [hep-th]]}.


\bibitem{Gaberdiel:2015mra} 
M.R.~Gaberdiel and R.~Gopakumar,
``Stringy Symmetries and the Higher Spin Square,''
J.\ Phys.\ A {\bf 48} (2015) 185402 
{\tt [arXiv:1501.07236 [hep-th]]}.


\bibitem{Gaberdiel:2015wpo}
M.R.~Gaberdiel and R.~Gopakumar,
``String Theory as a Higher Spin Theory,''
{\tt arXiv:1512.07237 [hep-th]}.


\bibitem{Baggio:2015jxa} 
M.~Baggio, M.R.~Gaberdiel and C.~Peng,
``Higher spins in the symmetric orbifold of K3,''
Phys.\ Rev.\ D {\bf 92} (2015) 026007
{\tt [arXiv:1504.00926 [hep-th]]}.

\bibitem{Gaberdiel:2015uca} 
M.R.~Gaberdiel, C.~Peng and I.G.~Zadeh,
``Higgsing the stringy higher spin symmetry,''
JHEP {\bf 1510} (2015)  101
{\tt [arXiv:1506.02045 [hep-th]]}.

\bibitem{Gaberdiel:2016xwo}
M.R.~Gaberdiel and M.~Kelm,
``The symmetric orbifold of N=2 minimal models,''
{\tt arXiv:1604.03964 [hep-th]}.

\bibitem{Jevicki:2015irq} 
A.~Jevicki and J.~Yoon,
``$S_N$ Orbifolds and String Interactions,''
J.\ Phys.\ A {\bf 49}  (2016) 205401
{\tt [arXiv:1511.07878 [hep-th]]}.

\bibitem{Gaberdiel:2014vca} 
M.R.~Gaberdiel and M.~Kelm,
``The continuous orbifold of $ \mathcal{N} = 2$ minimal model holography,''
JHEP {\bf 1408} (2014) 084 
{\tt [arXiv:1406.2345 [hep-th]]}.


\bibitem{FFJMM1}
B.~Feigin, E.~Feigin, M.~Jimbo, T.~Miwa and E.~Mukhin,
``Quantum Continuous $\mathfrak{gl}_\infty$: Semi-infinite construction of representations,''
Kyoto J. Math. 51, no. \textbf{2} (2011) 337 
{\tt [arXiv:1002.3100 [math.QA]] }.

\bibitem{FFJMM2}
B.~Feigin, E.~Feigin, M.~Jimbo, T.~Miwa and E.~Mukhin,
``Quantum Continuous $\mathfrak{gl}_\infty$: Tensor Products of Fock Modules and $\mathcal{W}_n$ Characters,''
Kyoto J. Math. 51, no. \textbf{2} (2011) 365 
{\tt [arXiv:1002.3113 [math.QA]] }.

\bibitem{FJMM1}
B.~Feigin, M.~Jimbo, T.~Miwa and E.~Mukhin,
``Quantum toroidal $\mathfrak{gl}_1$ algebra: plane partitions,''
Kyoto J. Math. 52, no. \textbf{3} (2012) 621 
{\tt [arXiv:1110.5310 [math.QA]] }.

\bibitem{Prochazka:2015deb}
T.~Proch\'{a}zka,
``W-symmetry, topological vertex and affine Yangian,''
{\tt arXiv:1512.07178 [hep-th]}.


\bibitem{Gaberdiel:2012ku}
M.R.~Gaberdiel and R.~Gopakumar,
``Triality in Minimal Model Holography,''
JHEP {\bf 1207} (2012) 127
{\tt [arXiv:1205.2472 [hep-th]]}.

\bibitem{GTL}
S.~Gautam and V.~Toledano-Laredo,
``Yangians and quantum loop algebras,"
Selecta Mathematica {\bf 19} (2013) 271
{\tt [arXiv:1012.3687 [math.QA]]}.

\bibitem{Tsym}
A.~Tsymbaliuk,
``The affine Yangian of $\mathfrak{gl}_1$ revisited,"
{\tt arXiv:1404.5240 [math.RT]}.

\bibitem{OhlssonSax:2011ms}
O.~Ohlsson Sax and B.~Stefanski,
``Integrability, spin-chains and the AdS3/CFT2 correspondence,''
JHEP {\bf 1108} (2011) 029
{\tt [arXiv:1106.2558 [hep-th]]}.

  
\bibitem{Sax:2012jv}
O.~Ohlsson Sax, B.~Stefanski, and A.~Torrielli,
``On the massless modes of the AdS3/CFT2 integrable systems,''
JHEP {\bf 1303} (2013) 109
{\tt [arXiv:1211.1952 [hep-th]]}.


\bibitem{Borsato:2015mma}
R.~Borsato, O.~Ohlsson Sax, A.~Sfondrini and B.~Stefański,
``The $\mathrm{AdS}_3\times \mathrm{S}^3\times \mathrm{S}^3\times\mathrm{S}^1$ worldsheet S matrix,''
J.\ Phys.\ A {\bf 48} (2015) no.41,  415401
{\tt [arXiv:1506.00218 [hep-th]]}.

\bibitem{Sfondrini:2014via}
A.~Sfondrini,
``Towards integrability for AdS3/CFT2,''
{\tt arXiv:1406.2971 [hep-th]}.


\bibitem{Bergshoeff:1989ns}
E.~Bergshoeff, M.P.~Blencowe and K.S.~Stelle,
``Area Preserving Diffeomorphisms and Higher Spin Algebra,''
Commun.\ Math.\ Phys.\  {\bf 128} (1990) 213.


\bibitem{Bergshoeff:1990yd}
E.~Bergshoeff, C.N.~Pope, L.J.~Romans, E.~Sezgin and X.~Shen,
``The Super $W$(infinity) Algebra,''
Phys.\ Lett.\ B {\bf 245} (1990) 447.


\bibitem{Depireux:1990df}
D.A.~Depireux,
``Fermionic realization of W(1+infinity),''
Phys.\ Lett.\ B {\bf 252} (1990) 586.


\bibitem{Bakas:1990ry} 
I.~Bakas and E.~Kiritsis,
``Bosonic realisation of a universal W algebra and $\mathbb{Z}_{\infty}$ parafermions,"
Nucl.\ Phys.\ B {\bf 343} (1990) 185
[Erratum ibid.\ B {\bf 350}, 512 (1991)].

\bibitem{Gaberdiel:2013jpa}
M.R.~Gaberdiel, K.~Jin, and W.~Li,
``Perturbations of ${\cal W}_\infty$ CFTs,''
JHEP {\bf 1310} (2013) 162
{\tt [arXiv:1307.4087 [hep-th]]}.
 

\bibitem{Gaberdiel:2011aa}
M.R.~Gaberdiel and P.~Suchanek,
``Limits of Minimal Models and Continuous Orbifolds,''
JHEP {\bf 1203} (2012) 104
{\tt [arXiv:1112.1708 [hep-th]]}.

\bibitem{combinatorics}
R.P.~Stanley,
``Enumerative combinatorics," vol 1.
2nd ed., Cambridge University Press (2012). 

\bibitem{Keane:2002}
J.~Keane, posted on the 
`On-line encyclopedia of integer sequences', 
{\tt https://oeis.org/A000712} (2002).

\bibitem{HardyRamanujan}
G.H.~Hardy and S.~Ramanujan, 
``Asymptotic Formulae in Combinatory Analysis," 
Proc.\ London Math.\ Soc.\ {\bf 17} (1918) 75.

\bibitem{Wright1931}
E.~Wright, 
``Asymptotic partition formulae I. Plane partitions,"
The Quarterly Journal of Mathematics, \textbf{1} (1931) 177.


\bibitem{BFM}
M.~Bershtein, B.~Feigin and G.~Merzon,
``Plane partitions with a `pit': generating functions and representation theory,''
{\tt  [arXiv:1512.08779 [math-CO]]}.

\bibitem{Gaberdiel:2011wb}
M.R.~Gaberdiel and T.~Hartman,
``Symmetries of Holographic Minimal Models,''
JHEP {\bf 1105} (2011) 031
{\tt [arXiv:1101.2910 [hep-th]]}.

\bibitem{Gaberdiel:2011zw}
M.R.~Gaberdiel, R.~Gopakumar, T.~Hartman, and S.~Raju, 
``Partition functions of holographic minimal models,''
JHEP {\bf 1108} (2011) 077
{\tt [arXiv:1106.1897 [hep-th]]}.

\bibitem{Perlmutter:2012ds}
E.~Perlmutter, T.~Prochazka, and J.~Raeymaekers,
``The semiclassical limit of W$_N$ CFTs and Vasiliev theory,''
JHEP {\bf 1305} (2013) 007
{\tt [arXiv:1210.8452 [hep-th]]}.


\bibitem{Castro:2011iw}
A.~Castro, R.~Gopakumar, M.~Gutperle, and J.~Raeymaekers,
``Conical defects in higher spin theories,''
JHEP {\bf 1202} (2012) 096
{\tt [arXiv:1111.3381 [hep-th]]}.


\bibitem{Bouwknegt:1992wg}
P.~Bouwknegt and K.~Schoutens,
``W symmetry in conformal field theory,''
Phys.\ Rept.\  {\bf 223} (1993) 183
{\tt [arXiv:hep-th/9210010]}.


\bibitem{sergeev}
A.N.~Sergeev,
``Representations of the Lie superalgebras gl(n,m) and Q(n) on the space of tensors,"
Funct.\ Anal.\ Appl.\ {\bf 18} (1984) 70.

\bibitem{berele}
A.~Berele and A.~Regev,
``Hook Young diagrams with applications to combinatorics and to representations of 
Lie superalgebras,"
Adv. Math. {\bf 64} (1987) 118.

\bibitem{Candu:2012jq}
C.~Candu and M.R.~Gaberdiel,
``Supersymmetric holography on $AdS_3$,''
JHEP {\bf 1309} (2013) 071
{\tt [arXiv:1203.1939 [hep-th]]}.



\end{thebibliography}

\end{document}